%Paper: astro-ph/9311066
%From: <fbernard@chipmunk.cita.utoronto.ca>
%Date: Thu, 25 Nov 93 16:54:50 EST

%The nonlinear evolution of rare events, Plain TeX
\magnification\magstep 1
\headline={\ifnum\pageno=1\hfil\else\hfil\tenrm--\ \folio\ --\hfil\fi}
\footline={\hfil}
\hsize=6.0truein
\vsize=8.54truein
\hoffset=0.25truein
\voffset=0.25truein
\baselineskip=12pt

\parskip=0.2 cm
\parindent=1cm
\tolerance=10000
\pretolerance=10000
\def\pp{\parshape 2 0truecm 15truecm 2truecm 13truecm}
\def\apjref#1;#2;#3;#4; {\par\pp#1, {#2}, {#3}, #4 \par}
\def\bookref#1;#2;#3; {\par\pp#1, { #2}, {\rm #3} \par}
\def\prepref#1;#2; {\par\pp#1, {#2} \par}
\def\ie{{\it i.e. }}
\def\eg{{\it e.g.}}
\def\etal{{\it et al. }}
\def\ltsima{$\; \buildrel < \over \sim \;$}
\def\simlt{\lower.5ex\hbox{\ltsima}}
\def\gtsima{$\; \buildrel > \over \sim \;$}
\def\simgt{\lower.5ex\hbox{\gtsima}}

\def\Om{\Omega}
\def\Th{\Theta}

\def\la{\lambda}

\def\rhob{\overline{\rho}}
\def\mg{\big <}
\def\md{\big >}
\def\mdi{{\big >}_{\delta_i}}
\def\mdL{{\big >}_{\delta_L}}
\def\sym{\hbox{cyc}}

\def\mM{{\cal M}}

\def\mG{{\cal G}}
\def\d{\hbox{d}}
\def\ii{\hbox{i}}
\def\dta{\delta}

\def\WTH{W_{\hbox{TH}}}

\def\WTHd{\WTH\left(\vert\vk_1+\vk_2\vert\ R_0\right)}
\def\WTHt{\WTH\left(\vert\vk_1+\vk_2+\vk_3\vert\ R_0\right)}
\def\DoDi{{D(t)\over D(t_i)}}

\def\vx{\hbox{\bf x}}
\def\vr{\hbox{\bf r}}
\def\vq{\hbox{\bf q}}

\def\vk{\hbox{\bf k}}
\def\vK{\hbox{\bf K}}
\def\gradx{\nabla_{\hbox{\rm x}}\,}
\def\gradq{\nabla_{\hbox{\rm q}}\,}
\def\dk{d_{\vk}}

\def\jk{j_{\vk}}
\def\Der{{\cal T}}
\def\Cexp{C_{\hbox{exp}}}
\def\Cscat{C_{\hbox{scat}}}
\def\un{^{(1)}}
\def\deux{^{(2)}}
\def\trois{^{(3)}}
\def\intk{\int{\d^3\vk\over(2\pi)^{3/2}}}

\def\intt{\int{\d^3\vk_1\over(2\pi)^{3/2}}
{\d^3\vk_2\over(2\pi)^{3/2}}{\d^3\vk_3\over(2\pi)^{3/2}}}

\vglue .5 truecm
\rightline{CITA 93/14}
\vglue 1.3 cm
\centerline{\bf THE NONLINEAR EVOLUTION}
\centerline{\bf OF RARE EVENTS}
\vskip .4 cm
\centerline{by}
\vskip .4 cm
\centerline{Francis BERNARDEAU}
\vskip .3 cm
\centerline{CITA, 60 St George St., Toronto, Ontario, Canada M5S 1A1}
\vskip 1. truecm
\centerline{\bf ABSTRACT}
In this paper I consider the {\sl nonlinear}
evolution of a rare density fluctuation
in a random density field with Gaussian fluctuations,
and I rigorously show that it
follows the spherical collapse dynamics applied to its mean initial
profile.
This result is valid
for any cosmological model and is independent of the shape of the
power spectrum. In the early stages of the dynamics the density contrast
of the fluctuation is seen to follow with a good accuracy the form
$$\dta=(1-\dta_L/1.5)^{-1.5}-1,$$
where $\dta_L$ is the linearly extrapolated overdensity.

I then investigate the validity domain of the rare event approximation
in terms of the parameter
$\nu=\dta_L/\sigma$ giving the initial overdensity scaled by the
rms fluctuation at the same mass scale, and find that it
depends critically on the shape of the power
spectrum. When the power law index $n$ is lower than $-1$ the
departure from the spherical collapse is expected to be small, at least
in the early stages of the dynamics, and even for moderate values
of $\nu$ ($\vert\nu\vert\ge 2$).
However, for $n\ge -1$ and whatever the value of $\nu$,
the dynamics seems to be  dominated by the small--scale fluctuations and
the subsequent
evolution of the peak may not be necessarily
correlated to its initial overdensity. I discuss
the implications of these results for the nonlinear dynamics
and the formation of astrophysical objects.
\vskip 1.5 cm
\leftline{\sl To appear in The Astrophysical Journal, May 10, 1994}
\vskip 1 truecm

 \vfill\eject

\centerline{\bf 1. INTRODUCTION}

The gravitational instability scenario is the most widely accepted
answer to the problem of the formation of the present structures of the
universe. In such a view the astrophysical objects such as
galaxies or clusters of galaxies
are thought
to be formed by the collapse of initial density fluctuations existing
in the density field after the recombination. If it is well known that
such perturbations are gravitationally unstable and are bound to grow
under the influence of gravity, the complete dynamics of a
density perturbation
embedded in a fluctuating density field is not understood.
Indeed it requires the resolution of nonlinear
equations which has not been done yet.
However
the behavior of particular objects for which the initial conditions are
well defined,
such as a perturbation with an initial top--hat density profile,
can be calculated up to their final collapse.
These results have given birth to some theoretical derivations of the
mass distribution function of the objects present in our universe
with the so called Press and Schechter formalism (Press \& Schechter 1974).
The main hypothesis on which this formalism is based is that any part
of the universe will eventually collapse and form a virialized object
according to its initial overdensity and within the time scale given
by the spherical collapse model (\eg, Peebles 1980).
It has also been suggested that it is more likely
that the objects form at the location of the peaks, maxima of the density
field (Kaiser 1984).
It leads to the peak formalism developed by Bardeen et al. (1986).
In any case, however, the time scale required for the formation of
a virialized object is thought to follow the spherical collapse model.
These formalisms are the only analytical models available to derive
the epoch of formation of astrophysical objects such as quasars
(Efstathiou \& Rees 1988), or
to derive the number density of objects like clusters
(Peebles, Daly \& Juszkiewicz 1989). Except with the use
of numerical simulations, this is the only way to constrain the various
cosmological models with such observations. It is thus of crucial interest
to know whether the spherical collapse is a good approximation for the
formation of isolated structures in a fluctuating density background.

Peebles (1990) recently reconsidered this old problem arguing that the
existence of small--scale fluctuations should slow down the dynamics
of the collapse. In his view the non--radial motions should initiate a
``previrialization'' process. He indeed presents some numerical
experiments in which such a trend  appears with a large magnitude.
These results may be at
variance with what has been previously obtained by Efstathiou et al. (1988)
and by Bond et al. (1990)
when they checked the validity of the Press and Schechter formula, and
also at variance with the results of Little, Weinberg \& Park (1991) and
Evrard \& Crone (1992) in which they
explicitly check the influence of a variable cut--off at small scale
in the power spectrum, that fails to detect any ``previrialization''
effect.

In this paper I consider this problem from an analytical point of view
by analysing the early stages of the dynamics of a constrained
Gaussian random field
with perturbation theories. This
paper, however, is not devoted to a new perturbative theory
but aims to consistently calculate some statistical quantities describing
the dynamics of a given density fluctuation.
I focus my interest on the evolution
of the volume of such a rare fluctuation during its nonlinear evolution.
The volume is simply defined as the volume occupied at any time by the
particles that were in the initial density perturbation.
Although it is certainly a na\"\i ve picture for the
formation of an object that likely involves accretion or ejection
of matter, such an approximation is in the spirit of the current
theories of the
mass distribution function for which it is assumed that the matter content
of a fluctuation is conserved during the collapse. In the discussion
I will comment on the relevance of this hypothesis.

In part 2, I compute the expectation value of the size of a density
perturbation in the limit of rare events.
It is shown that in such a limit it can be derived even in
the fully nonlinear regime and it then follows the spherical collapse
solution.
Part 3 is devoted to the discussion of the
accuracy of this limit, that is, the first corrections
expected from this limit case and their magnitudes.
The case of a realistic model for the structure formation, \ie
a CDM model, is considered.
A large fraction of the mathematical content of the resolution of this problem
is given in appendices. The first one is devoted to
usual results on the expectation values of non--Gaussian variables,
and the second to the computation of two integrals involving the
top--hat filter function. The third one is the most important. It contains
the rigorous demonstration of the results presented in part 2.
The last one is devoted to tedious calculations of interest for the discussion
presented in part 3.

 \vskip 1 cm

\centerline{\bf 2. THE NONLINEAR EVOLUTION OF RARE EVENTS}
\vskip .5 cm
\leftline{2.1 \sl The initial conditions}

I assume that the structures of the universe form from gravitational
instabilities created by Gaussian fluctuations in a pressureless matter
density field.
The overdensity
field $\delta(\vr)$ is then a random field so that its Fourier transform
components $\delta_{\vk}$ defined by,
$$\delta(\vr)=\int {\d^3\vk\over (2\pi)^{3/2}}
\ \delta_{\vk}\ e^{\ii\vk. \vr}\,,\eqno(1)$$
are random complex variables with independent phases.
The reality constraint of $\dta(\vr)$ implies
$$\mg\dta_{\vk_1}\dta_{\vk_2}\md=\dta_3(\vk_1+\vk_2)P(k_1)\eqno(2)$$
where $\dta_3(\vk)$ is the three
dimensional Dirac distribution.
The power spectrum, $P(k)$, determines the full process of the structure
formation.

Throughout this paper I assume that within a spherical volume
of radius $R_0$ there is a given initial overdensity $\dta_i$,
$$\dta_i={1\over V_0}\int_{V_0} \d^3 \vr\ \dta(\vr).\eqno(3)$$
If the constraint (3) is fulfilled I will say that there is a
``peak'' in the given volume with the initial overdensity $\dta_i$.
This is a rather poor definition that makes sense at least when $\dta_i$
is large in a sense that will be given afterwards.
One could have thought of stronger constraints for a given
location to be called a peak, and, following Bardeen et al. (1986),
required that this volume is at a maximum of the density field.
Such a requirement will not change the qualitative results presented
throughout this paper as it will be discussed at the end.
Moreover the results presented in the following are valid for
positive values of $\dta_i$ as well as negative values. So it will
describe also the nonlinear evolution of rare voids.

Once a peak is supposed to lie at a given location
the resulting constrained random field is not only a function
of $P(k)$ but also a function of $\dta_i$.
This quantity is a Gaussian variable related
to the $\dta_{\vk}$. Its variance is given by
 $$\eqalign{
\sigma_i^2=\mg\dta_i^2\md&
={1\over V_0^2}\int_{V_0} \d^3 \vr\ \d^3 \vr'\mg \dta(\vr)\dta(\vr')\md\cr
&={1\over V_0^2}
\int_{V_0}\d^3 \vr \d^3\vr' \int
{\d^3\vk\over (2 \pi)^{3/2}}\,{\d^3\vk'\over (2 \pi)^{3/2}}\,
\mg\dta_{\vk}\dta_{\vk'}\md e^{\ii\vk.\vr+\ii\vk'.\vr'}.\cr}\eqno(4)$$
For convenience I introduce the Fourier transform of the top--hat window
function,
$${1\over V_0} \int_{V_0}\d^3\vr\,e^{\ii\vk.\vr}={3\over (kR_0)^3}
\left[\sin(kR_0)-kR_0\cos(kR_0)\right]\equiv \WTH(kR_0).\eqno(5)$$
The resulting variance for the initial overdensity is
$$\sigma_i^2=\int{\d^3\vk\over (2\pi)^3}\ \WTH^2(k\,R_0)\,P(k).\eqno(6)$$

The purpose of this paper is to study
the expected dynamics of the random field once the constraint (3) has
been set. One obvious and
direct thing to do is to consider the expectation values of the Fourier
components of the random field.
In the following I denote $\mg.\mdi$ the expectation value of any
random quantity when the constraint (3) is set. A priori any such
quantity depends both on the power spectrum and on the value of $\dta_i$.
The variables $\dta_i$ and $\dta_{\vk}$ are both Gaussian and they are
correlated together, so that
$$\eqalign{
\mg\dta_{\vk}\mdi&={\mg\dta_{\vk}\dta_i\md\over\mg\dta_i^2\md}\dta_i\cr
&={P(k)\,\WTH(k\,R_0)\over (2\pi)^{3/2}\ \sigma_i^2}\ \dta_i.\cr}\eqno(7)$$

The mean initial field exhibits a symmetric density perturbation, the profile
of which is determined by the shape of the power spectrum. The expected
local overdensity within a distance $r$ from the center of the perturbation
is
$$\mg\dta(<r)\mdi={\int{\d^3\vk/ (2\pi)^3}\ P(k)\WTH(k\,R_0)\WTH(k\,r)
\over \sigma_i^2}\ \dta_i\eqno(8)$$
and the local overdensity at a distance $r$ from the center can be easily
derived from this formula.
The nonlinear evolution of a spherically symmetric perturbation
with this mean density profile is also
straightforward to compute:
the trajectory of the matter at a distance $r$ of the center
depends only on the matter contained inside the radius $r$.
The evolution of the profile of the object is shown in Fig. 1
as a function of
time for various initial power spectra. There are a priori no reasons for the
nonlinear evolution of these mean profiles to reproduce the mean nonlinear
profiles. However,
for a very dense perturbation one would expect that the dynamics of the
collapse somehow follows this behavior. This is
the central problem addressed in this paper.

\vskip .5cm
\leftline{2.2 \sl The equations of motion}

In the nonlinear evolution of the objects with the previous profiles,
the size of the perturbation
turns out to be the only quantity that is independent of the shape of
the power spectrum. This is the reason why,
also in the nonlinear constrained dynamics,
I focus my interest on it.
The size of the perturbation is assumed to be simply given by the
volume occupied by the matter that was
initially inside the radius $R_0$ all along the nonlinear evolution.
A Lagrangian calculation is the most
adequate approach for the purpose of this calculation.
In the initial unperturbed field, fluctuations are assumed to create
initial random displacements that will be amplified by the gravitational
dynamics. The unperturbed comoving positions are denoted $\vq$, and the exact
comoving positions, $\vx$, are related to $\vq$ through a displacement
field $\Psi(t,\vq)$,
$$\vx=\vq+\Psi(t,\vq).\eqno(9)$$

The volume $V(t)$ subsequently occupied by the particles in the real
space is related to the displacement field through space derivatives. Indeed
it reads
$$V(t)=\int_{V}\d^3\vx=\int_{V_0}\left\vert{\partial \vx\over\partial\vq}
\right\vert\,\d^3\vq.\eqno(10)$$
The jacobian of the transformation $J(t,\vq)$ defined by
$$J(t,\vq)=\left\vert{\partial \vx\over\partial\vq}\right\vert\eqno(11)$$
is then the quantity of interest for the calculation of the
volume. It is the determinant of the
matrix the elements of which are
$${\partial x_i\over \partial q_j}=\dta_{ij}+{\partial \Psi_i \over
\partial q_j}=\dta_{ij}+\Psi_{i,j}\eqno(12)$$
where $\dta_{ij}$ is the Kronecker symbol.
The motion equation of the point being at the position $\vx$
leads straightforwardly with the Poisson equation to
$$\gradx.\ddot{\vx}+2{\dot{a}\over a}\gradx.\dot{\vx}=-4\pi G
(\rho(\vx)-\rhob).\eqno(13)$$
The density $\rho(\vx)$ at the position $\vx$ is given by the matter
conservation constraint
$$\vert J(\vx)\vert\rho(\vx)=\rhob.\eqno(14)$$
In the following I assume that $J(\vx)$ is positive so that
$\vert J(\vx)\vert=J(\vx)$. This assumption is valid in the quasilinear
regime since $J(\vx)$ is then close to 1, but breaks after the
first shell crossing where $J(\vx)$ is zero. With this assumption we then
have
$$J(t,\vq)\left(\gradx.\ddot{\vx}+2{\dot{a}\over a}\gradx.\dot{\vx}\right)
=-4\pi\ G\ \rhob\left[1-J(t,\vq)\right].\eqno(15)$$
We assume that the time $t_i$ at which the initial fluctuations start to
grow is arbitrarily close to the origin. The decaying modes of any kind
are then subsequently erased so that the only purely growing modes
remain at the present time. This is likely to be an unjustified
approximation for the study of the virialization processes but we are only
interested here in the collapse prior to the virialization. The time
dependence of the growing mode in the linear approximation
will be denoted $D(t)$. It is just proportional to the expansion
factor in case of an Einstein--de Sitter universe but in general depends
on the cosmological parameters (see Lahav et al. 1991 for a review).
Another consequence of crucial interest is that the rotational
modes are suppressed, since they are simply expected to be diluted
with the expansion (\eg, Peebles 1980).
I then assume that the displacement field
is non--rotational, that is
$$\gradx\times\dot{\Psi}(\vx)=0.\eqno(16)$$

The resolution of the dynamics
in such an approach is entirely driven by the behavior
of the displacement field, so that the motion equations have to be
written only in term of $\Psi(t,\vq)$ and its derivatives.
The equation (11) then reads
$$\eqalign{
J(t,\vq)=&1+\gradq.\Psi+{1\over 2}\left[\left(\gradq.\Psi\right)^2-
\sum_{ij}\Psi_{i,j}\Psi_{j,i}\right]\cr
&+{1\over 6}
\left[\left(\gradq.\Psi\right)^3-3\gradq.\Psi\sum_{ij}\Psi_{i,j}\Psi_{j,i}
+2 \sum_{ijk}\Psi_{i,j}\Psi_{j,k}\Psi_{k,i}\right].\cr}\eqno(17)$$
where $\gradq$ is the gradient taken relatively to $\vq$
and the summations are made over the three spatial components. The
equation (15) involves a time derivative operator that I denote
$\Der(.)$ and that is defined by
$$\Der(A)=\ddot{A}+2{\dot{a}\over a}\dot{A}.\eqno(18)$$
Expressed in term of the displacement field only the equation (15)
reads
$$\eqalign{
\Der\left(\gradq.\Psi\right)+
\gradq.\Psi\,\Der\left(\gradq.\Psi\right)-&
\sum_{ij}\Psi_{i,j}\Der\left(\Psi_{j,i}\right)
+{1\over 2}\left(\gradq.\Psi\right)^2\Der\left(\gradq.\Psi\right)-\cr
-{1\over2}\Der\left(\gradq.\Psi\right)\sum_{ij}\left[\Psi_{i,j}\Psi_{j,i}
\right.-&\left.\gradq.\Psi\,\Der\left(\Psi_{i,j}\right)\Psi_{j,i}\right]
+\sum_{ijk}\Psi_{i,j}\Psi_{j,k}\Der\left(\Psi_{k,i}\right)\cr
&\ \ \ =-4\pi G\rhob\left[1-J(t,\vq)\right].\cr}
\eqno(19)$$
The equations (17) and (19) do not allow a complete resolution
of the dynamics. However, the displacement field is fully determined as soon
as the non--rotational constraint (16) has been set.
The resolution of these equations would lead to a complete knowledge of the
nonlinear dynamics (before shell crossings).
I am interested here in a particular quantity, \ie the
volume occupied by a peak all along its nonlinear evolution.

\vskip .5 cm
\leftline{\sl 2.3 The expectation value of the volume}

 The volume of the object is obviously a non--Gaussian random
variable and a complicated function of the random variables $\dta_{\vk}$,
which are correlated to the initial linear overdensity $\dta_i$. In this part
I am interested in the expectation value of the volume, $\mg V(t)\mdi$,
for a random density field that follows the constraint (3). This quantity
is a priori a function of $V_0$, of the power spectrum $P(k)$,
of the initial overdensity $\dta_i$ and of time.
Once the time and the shape of the power spectrum have been specified
this is a function of $\dta_i$ and $\sigma_i$ that contains the
normalization of the initial fluctuations. Equivalently it is a function
of the linearly extrapolated overdensity $\dta_L$,
$$\dta_L=\DoDi\dta_i,\eqno(20)$$
where $D(t)$ is the time dependence of the growing mode, and the
linearly extrapolated rms fluctuations at the same scale,
$$\sigma_L=\DoDi\sigma_i.\eqno(21)$$
The regime that is investigated concerns the collapse
of rare peaks  or the expansion of rare troughs before the first
shell crossing. As a result the linear
overdensity of the object, $\dta_L$, is at most of the
order of a few units but the rms
fluctuation at the same scale, $\sigma_L$, is small. The
rare peak limit then implies that the expectation value of $V(t)$
has to be taken for a vanishing value of $\sigma_L$ and a fixed value
of $\dta_L$.

Following the appendix A, such expectation value is given by
(Eqs. A.13, A.16)
$$\eqalign{\mg V(t)\mdL=&
\int_{-\infty}^{+\infty}\d z\,\mG_{V}(t,\sigma_L,\dta_L+\ii z)\,
\exp\left[-{z^2\over 2\ \sigma_L^2}\right]
/\int_{-\infty}^{+\infty}\d z\,
\exp\left[-{z^2\over 2\ \sigma_L^2}\right]\cr}\eqno(22)$$
with
$$\mG_{V}(t,\sigma_L,y)=\sum_{p=0}^{\infty}
{\mg V(t) \dta_L^p\md_c\over \sigma_L^{2p}}{y^p\over p!},\eqno(23)$$
since $\dta_L$ is a Gaussian variable.
The mean volume is then a complicated function of $\dta_L$ and
$\sigma_L$ which I want to calculate in the limit $\sigma_L\to 0$.
The leading term in the numerator of the ratio (22) is given by a saddle
point approximation (around $z\approx0$). The relation (22) then reads
$$\mg V(t)\mdL\approx \mG_V(t,\sigma_L=0,\dta_L).\eqno(24)$$
I would like to stress that
this expression is exact in the rare peak limit but involves
all orders of the perturbation theory through the function
$\mG_V(t,\sigma_L=0,\dta_L)$ (but assuming no shell crossing and
irrotationality). If the rare peak approximation
were released, the expression (24) would no longer be valid.
The part 3 is devoted to an analysis of the accuracy of this limit,
its real meaning, and its validity domain.

Each term involved in the expression of $\mG_V(t,\sigma_L,\dta_L)$
has then to be calculated when $\sigma_L=0$ (as we will see each term
is actually finite in spite of the denominators appearing in Eq. (23)).
The quantities of interest $\mg V(t) \dta_L^p\md_c$ have then to be calculated
at the lowest order in $\sigma_L$ (here $\dta_L$ is not a fixed parameter
but a Gaussian random variable).

The random variable $V(t)$ can be expanded relatively to $\dta_{\vk}$,
$$V(t)=\sum_{p=0}^{\infty} V^{(p)}(t)\eqno$$
where $V^{(p)}(t)$ contains exactly $p$ factors of  $\dta_{\vk}$
in its expression. The magnitude of $V^{(p)}(t)$ is
$$V^{(p)}(t)\sim V_0\,\delta_L^p.\eqno(25)$$
The order 0 is simply the comoving volume $V_0$;
the order 1 corresponds to the linear approximation: $V^{(1)}=-V_0\,
\dta_L$; and so on.
The cumulant, $\mg V(t) \dta_L^p\md_c$,
 appearing in the expression of $\mG_V$ (see appendix A for
a definition) connects $p+1$ random variables and then requires the product
of at least $2p$ random Gaussian variables. As a result the leading
contribution
to this cumulant when the rms fluctuation is small
comes from the order $p$ of $V(t)$. These arguments are the same
that lead to the scaling behavior of the $p$ order cumulants
of the distribution function of the density at large scale
(Fry 1986, Bernardeau 1992). The leading  contribution at small $\sigma_L$
is the exact result in the rare peak approximation, so that in such a case
we get
$$j_p(t)\equiv\mg V(t) \dta_L^p\md_c/V_0
= \mg V^{(p)}(t)\dta_L^p\md_c/V_0\eqno(26)$$
and this quantity is of the order of
$$j_p(t)\sim \sigma_i^{2p}.\eqno(27)$$
The function $\mG_V(t,\sigma_L=0,\dta_L)$ is then a finite
number for finite values of $\delta_L$.

At this stage the most difficult part
is the derivation of the function $\mG_V(t,0,y)$ (denoted
$\mG_V(t,y)$ in the following).
It cannot be done by a simple perturbative calculation since
all orders of the dynamics are involved. As it is a rather complicated and
long demonstration I present it only as an appendix (appendix C).
It contains absolutely no other approximations than the ones presented
previously and it gives the values of the whole series of $j_p(t)$
as written  in the equation (26).

The result of the appendix C is quite simple
and shows that the function $\mG_V(t,y)$ follows the differential equation,
$${\d^2\over\d t^2}\left(R_0 \left[\mG_V(t,y)/V_0\right]^{1/3}\right)=
{-(G\ \rhob+\Lambda/4\pi)\ V_0\over \left(R_0
\left[\mG_V(t,y)/V_0\right]^{1/3}\right)^2},
\eqno(28)$$
which is the motion equation describing a spherical collapse,
where $\left(R_0 \left[\mG_V(t,y)/V_0\right]^{1/3}\right)$ is the size
of the object ($\Lambda$ is a possible cosmological constant).
The behavior of $\mG_V(t,y)$ for small values of $y$,
($\mG_V(t,y)=V_0[1-y+\dots]$)
ensures that the corresponding overdensity is simply $y$.
The function $\mG_V(t,\dta_L)$ that gives the expectation volume
of a peak of overdensity $\dta_L$
is then simply the volume occupied at the time $t$
by an object having the same initial overdensity and
initial comoving volume, according to the spherical collapse
model. This result is independent of the
cosmological parameters ($\Omega$ and $\Lambda$). As it can be noticed,
the dependence on the power spectrum completely vanished and
an individual collapse is expected to follow
the same dynamics whatever the shape of the power spectrum. This
result, however, is not robust, as  it is discussed in part~3.
It is quite comforting to see that
the spherical collapse has something to do with the collapse of an
object embedded in a fluctuating universe.

\vskip .5 cm
\leftline{\sl 2.4 The nonlinear evolution of the profile, of the density}

The previous calculation deals with the expectation value of the volume
$V$ occupied by the particles that were in the initial volume $V_0$.
One obviously could have considered the particles that were initially
within the radius $\lambda R_0$ (whatever $\lambda$) and derive the
expectation value of the volume occupied by these particles. The same kind of
calculations can be done for this more general problem and the result
is that they occupy the volume
${\lambda^3\,V_0\,\mG_V[t,\mg\dta(<\lambda R_0)\mdL]}$ which is the exact
nonlinear evolution of the mean profile (this result is not a consequence of
the previous one but its demonstration is nearly identical).
The result is also valid for
underdense areas so that the spherical model solution is expected to
describe as well the collapse of rare peaks as the expansion
of rare troughs.

I propose a simple analytical expression
for the evolution of the volume valid as long as the density contrast
is not too large:
$$\mg V(t)\mdL=V_0\,\left[1-{\dta_L\over 1.5}\right]^{1.5},\eqno(29a)$$
where $\dta_L$ is the linear extrapolation of the initial
overdensity (eq. [20]). The dependence with the cosmological
parameters is entirely contained in the time dependence of $D(t)$.
The form (29a) is exact for $\Omega\to0$, $\Lambda=0$ and is a good fit
for any values of $\Omega$ and $\Lambda$, especially for the underdense
regions ($\dta_L<0$). It fails, however,
to reproduce with a good accuracy the position of the
singularity (Fig. 2).

The mean overdensity of the object,
simply given by $\mg V_0/V(t)\mdL-1$, can also
be calculated in the rare event limit. It involves a similar generating
function $\mG_{\dta}(t,y)$ that can be calculated in the rare event limit
(appendix C). The result in such a case is quite simple and reads,
$\mG_{\dta}(t,y)=V_0/\mG_{V}(t,y)-1$, so that the mean density is simply given
by the inverse of the mean volume and one can use the relationship,
$$\mg \dta\mdL=\left[1-{\dta_L\over 1.5}\right]^{-1.5}-1.\eqno(29b)$$

As a result it is shown that in the
rare event limit the density fluctuations behave as if they had the mean
initial profile described in the section 2.1. In particular the density
fluctuations are expected to be spherical (otherwise the timescale of the
collapse would not be the one of the spherical collapse)
and the outskirts of clusters
for instance are well described by such an approach (while the inner part is
subject to processes that are beyond the scope of these calculations).
This result may not be surprising. It is simply that the rare event constraint
of the random Gaussian field (3) is strong enough to determine the shape of
the initial profile and its subsequent dynamics. What has been obtained here
is a rigourous demonstration that in the rare event limit, in a field
with Gaussian fluctuations, the dynamics of a constained field
follows the spherical collapse
up to the first singularity (that is at $\dta_L\approx 1.68$ for an
Einstein--de Sitter universe). The subsequent evolution cannot be studied
with a perturbation theory (the reason is that the determinant
$J(\vx)$ is not necessarily positive and taking the absolute value
cannot be obtained by means of a perturbative expansion).

In practice however, the astrophysical objects may approach such a limit
but never reach it exactly. This is the problem addressed in the
third part of this work.

 \vskip 1 cm

\centerline{\bf 3. THE ACCURACY OF THE RARE EVENT LIMIT}

The interest of the previous calculations is not only that we recover a known
dynamics but that it is now viewed as a limit process. The accuracy
of this limit can now be checked for each step of the calculation.
It can be evaluated either by the departure
of the exact mean collapse from the spherical collapse, or by
the magnitude of the scattering around such a behavior.
Firstly I examine the first corrections to the spherical collapse
that appear when the rare event limit is released.
The parameter that is supposed to be large is defined by
$$\nu={\dta_L\over \sigma_L}.\eqno(30)$$
The probability distribution of $\nu$ is normal and rare events
correspond to large values of $\nu$, either positive or negative.
In case of clusters for instance the values for $\nu$
derived from the observations are of the order of 2 to 4
(Peebles et al. 1989). For quasars
the values required for $\nu$ may be greater. In the following I
will then be concerned by the values of $\nu$ of this order of
magnitude or greater.

\vskip .5 cm
\leftline{3.1 \sl The corrections to the mean collapse}

The mean value of the volume can be expanded with respect to
$\sigma_L$,
$$\mg V\mdL=\mg V\mdL^{(0)}+\sigma_L^2\ \mg V\mdL^{(2)}+\dots$$
The term $\mg V\mdL^{(0)}$ is the one that has been calculated
previously in the rare event limit. The first correction is given by
the term $\sigma_L^2 \mg V\mdL^{(2)}$ (the order 1 in $\sigma_L$
is identically zero). To discuss the results in term of $\nu$
this correction can be written as
${\dta_L^2/ \nu^2}\mg V\mdL^{(2)}$ which due to this
change of variable gives a correction starting at the quadratic
order in $\dta_L$. The evaluation of $\mg V\mdL^{(2)}$ has to be done
by a perturbative calculation around the rare event limit, that
is around $\sigma_L=0$.
During the calculations of part~2  the $\sigma_L=0$ limit
has been invoked two times, once
to derive the expression of the mean value of $V(t)$ as given by (22), then
to calculate $j_p$, in Eq. (26).
It would be possible to calculate (22) exactly but
the derivation of the parameters $j_p$ is
more complicated and cannot be done in general. Actually it will not be
possible to derive $\mg V\mdL^{(2)}$ exactly but only the first
two terms of its expansion with respect to $\dta_L$. These terms are
partly given by the first correction in $\sigma_L^2$ of $j_0$ and $j_1$.
Indeed in general the value of $j_p$ can be seen as an expansion with respect
to $\dta_{\vk}$,
$$V_0\,j_p(t)
=\mg V^{(p)}(t)\dta_L^p\md_c+\mg V^{(p+2)}(t)\dta_L^p\md_c+...\eqno(31)
$$

The first term of this expansion is the one already calculated, and it has
been shown to correspond to the spherical collapse. The corrective terms
cannot be calculated as easily and the calculation has to be made
by hand for each of the terms.
For $p=0$ the situation is extremely simple since the corrections
are all identically zero. It just means that the ensemble average of the
volume simply
follows the comoving volume. The first non--trivial term appears for
$p=1$.
The leading term is simply the one coming from the linear approximation
and the first correction is then
$$j_1(t)=-\sigma_L^2
+{1\over V_0}\mg V^{(3)}(t)\dta_L\md_c+...\eqno(32)$$
 The calculation of this corrective term is presented in the appendix D
in the case of an Einstein--de Sitter universe.
According to the relation (D.19) the expression
of $j_1$ is then
$$\eqalign{
j_1(t)=-\sigma_L^2\ &\left[
1-{4\sigma_L^2\over 21\sigma_i^2}\int {\d^3 \vk \over (2 \pi)^{3}}
\WTH^2(k\,R_0) P(k)\int {\d^3 \vk' \over (2 \pi)^{3}} P(k') f(k'/k)\right]\cr
=-\sigma_L^2\ &\left[
1-\sigma_L^2\ \Cexp\right]\cr}\eqno(33)$$
where the corrective coefficient $\Cexp$ is given by
$$\Cexp={4
\int {\d^3 \vk \over (2 \pi)^{3}} \WTH^2(k\,R_0) P(k)
\int {\d^3 \vk' \over (2 \pi)^{3}} P(k') f(k'/k)
\over 21
\left[\int{\d^3\vk\over (2\pi)^3}\ \WTH^2(kR_0)\,P(k)\right]^2}
\eqno(34)$$
and $f(\tau)$ is the function defined in (D.20). It is plotted in
Fig. 3. The corrective term appears to be the product of one
non--dimensional parameter, $\Cexp$, and a time dependent factor
which is simply the linear extrapolated
value of the variance at the scale $R_0$ for the present time.
This allows one to derive the exact form of
$\mg V(t)\mdL$ up to the third order in $\dta_L$ as a function of
$\nu$.
The results are presented for an Einstein--de Sitter universe.
In such a case we have
$$\eqalign{
j_2(t)&={8\over 21}\sigma_i^4 ,\ \ \
j_3(t)={20\over 189}\sigma_i^6
.\cr}\eqno(35)$$
The resulting form for the expectation value of the volume when
the corrections coming from the exact
calculation of the integral (22) are included, is
$$\mg V(t)\mdL=V_0\left[
1-\dta_L+{4\over 21}
(1-1/\nu^2)\,\dta_L^2
+{10\over 567}
\left[1+{-3+567/10\ \Cexp\over \nu^2}\right]\,\dta_L^3+..\right].\eqno(36)$$
The dynamics
of the spherical collapse is recovered in the limit $\nu\to\infty$.
The accuracy of this limit then depends on the values
taken by $\Cexp$. As can be seen from the equation (34) the value of
this coefficient is a function of the shape of the power spectrum.
In case $P(k)$ has a power--law shape
$$P(k)\propto k^n\eqno(37)$$
the resulting values for $\Cexp$ are given in Fig. 4a. The most
remarkable feature is that the coefficient diverges for $n\ge-1$.
This divergence comes from the large values of $k$ and is due
to the shape of the
function $f(\tau)$ for $\tau\to\infty$ ($f(\tau)\sim 5/8\tau^2$ when
$\tau\to\infty$). The small
scale fluctuations are then responsible for the infinite value
of $\Cexp$ when $n\ge-1$, and for such power spectra
the rare event limit can never be reached, {\sl whatever the value of $\nu$}.
In order to check the importance of this divergence, I also
considered the case of a power spectrum with a cut--off
$k_c$ for large values of $k$. I then give in Fig. 4b the behavior of
$\Cexp$ when the position of the cut--off varies compared to the inverse
of the size of the fluctuation. The results are given for various
values of $n$. It appears that for $n=-1$, which is the critical case,
the divergence is only logarithmic so that a very large dynamical range
is required for such an effect to be noticed. For $n=0$ the divergence
is more rapid (as $[R_0k_c]^2$)
and such an effect could be seen in numerical simulations.

When $n<-1$ the corrective coefficient is small and the values of
$\nu$ of a few units are enough to ensure a behavior close to the
spherical case. For $n=-2$ we have $\Cexp\approx0.11$.
In Fig. 5, I present the expected density of
the object (the mere inverse of the volume) as a function of
the initial overdensity for $\nu=4$ and $\nu=2$. The curves
correspond to
$$\mg V_0/V(t)\mdL=
1+\dta_L+{17\over 21}
(1+{4\over 17\nu^2})\,\dta_L^2
+{341\over 567}
\left[1+{246/341-567/341\ \Cexp\over \nu^2}\right]\,\dta_L^3
+...\eqno(38)$$
The relation (38) has not been obtained as a mere expansion of the
inverse of the relation (36) but as an exact determination of
$\mg V_0/V(t)\mdL$ up to the third order. It happens that it is the same.
The departures from the
spherical collapse (at the same order) remain small.

For the known astrophysical objects this is then of crucial interest
to know the value of $n$ at their mass scale. For galaxies and quasars
a direct observation is not possible since the universe is now
fully nonlinear at their mass scale, but according to most of the
theories based on the gravitational instability scenario, $n$ is likely
to be of the order of $-2$ or smaller. For the clusters the situation is more
critical. Direct measurements of the mass fluctuations through galaxy counts
give $n\approx-1.3$ according to Peacock (1991) for the APM angular survey,
$n\approx -1.4$ according to Fisher et al. (1993) for the IRAS galaxy survey.
The use of the temperature distribution of the clusters allows one (with
the Press  and Schechter formalism) to do a direct estimation of the shape
of $P(k)$ for the matter field. It leads, according to Henry \& Arnaud
to $n\approx -(1.7^{+0.65}_{-0.35})$ and according to Oukbir \& Blanchard
(1992) to $n\approx -2$ independently of the density of the universe.
It then seems that for the clusters it is also fair
to use the spherical collapse model.

The case of the CDM model is considered in Fig. 6. The coefficient $\Cexp$
is computed with
$$P(k)\propto {k\over
(1+\alpha\,k+\beta\,k^{3/2}+\gamma\,k^2)^2},\eqno(39)$$
$\alpha=1.71l$, $\beta=9.0l^{3/2}$, $\gamma=1.0l^2$ and $l=4.0$
corresponding to $h=0.5$, $\Omega=1.$ (Davis et al. 1985). The corrective
coefficient remains essentially finite for the masses of interest, so that
the spherical collapse model is expected to hold for such a shape of power
spectra.

\vskip .5 cm
\leftline{\sl 3.2 Fluctuations around the mean behavior}

 The purpose of this part is to derive the value of
$\mg V^2(t)\mdL-\mg V(t)\mdL^2$ to evaluate the magnitude of the scattering
around the mean collapse.

First of all we can calculate this expression in the same
limit, \ie the rare event approximation. According to the appendix
A, this function is given by
$$\eqalign{\mg V^2(t)\mdL=&
\int_{-\infty}^{+\infty}\d z\,
\left[\mG_{V}^2(t,\dta_L+\ii z)+\mG_{V^2}(t,\dta_L+\ii z)\right]
\exp\left[-{z^2\over 2\ \sigma_L^2}\right]\cr
\ & /\int_{-\infty}^{+\infty}\d z\,
\exp\left[-{z^2\over 2\ \sigma_L^2}\right]\cr}\eqno(40)$$
with
$$\mG_{V^2}(t,y)=\sum_{p=0}^{\infty}
{\mg V^2(t) \dta_L^p\md_c\over \sigma_L^{2p}}{y^p\over p!}.\eqno$$
In the rare event limit, and for the same reason as before we have,
$$\mg V^2(t) \dta_L^p\md_c\approx
\sum_{r=1}^{p+1}\mg V^{(r)}(t)\ V^{(p+2-r)}(t)\ \dta_L^p\md_c.\eqno(41)$$
The latter expression can be calculated exactly. The result given in the
appendix C is quite simple and reads,
$$\mG_{V^2}(t,y)\approx\sigma_L^2
\left[{\d\over\d y}\mG_V(t,y)\right]^2.\eqno(42)$$
This result is only valid in the rare event approximation: the function
$\mG_V$ that appears in this expression corresponds to the spherical collapse
and the first corrections to (42) are of the order of $\sigma_L^4$.
The calculation of the expression (40) has then to be made in a consistent
way. We obtain
$$\eqalign{
\mg V^2(t)\mdL=&\mG_V^2(t,\dta_L)
+\sigma_L^2\mG_{V^2}(t,\dta_L)\cr
&\ -\sigma_L^2\left(\mG_V(t,\dta_L){\d^2\over\d y^2}\mG_V(t,\dta_L)+
\left[{\d\over\d y}\mG_V(t,\dta_L)\right]^2\right)+\dots\cr}\eqno(43)$$
In the first term $\mG_V$ does not only correspond to the spherical
collapse but also contains corrections of the order of $\sigma_L^2$
when the approximation (24) is released. The last
term corresponds to the first correction of (40) around the saddle
point position (obtained by an expansion of $\mG_V^2(t,y)$ around $z=0$).
The first neglected terms are of the order of $\sigma_L^4$.
At the same order we have
$$\mg V(t)\mdL^2=\mG_V^2(t,\dta_L)-\sigma_L^2\mG_V(t,\dta_L)
{\d^2\over\d y^2}\mG_V(t,\dta_L)+\dots\eqno(44)$$

The equations (42-44) imply that all the terms that are
independent of $\sigma_L$
or proportional to $\sigma_L^2$ vanish, leading to a scattering of the form
of,
$$\left[\mg V^2(t)\mdL-\mg V(t)\mdL^2\right]^{1/2}\sim
\sigma_L^2\ \Cscat(t,\dta_L)\ V_0,\eqno(45)$$
at the lowest order in $\sigma_L$. The coefficient $\Cscat(t,\dta_L)$
is a finite function, which
in case of an Einstein--de Sitter universe is a function of
$\dta_L$ only. This result just means that there is no scattering when
the rare event approximation is valid: the magnitude of the scattering is just
equal to the magnitude of the error between the spherical collapse model
and the exact mean behavior.

To be more convincing I derive the value of $\Cscat$ for $\dta_L=0$.
A careful examination of the various contributions leads to the expression,
$$\Cscat(\dta_L=0)=
{1\over V_0^2\sigma_L^2}
\left[\mg \left(V\deux(t)\right)^2\md-{1\over2}
{\mg V\deux(t)\,\dta_L^2\md^2\over \sigma_L^4}\right]^{1/2}.\eqno(46)$$

It is calculated in the appendix D and plotted in Fig. 7
as a function of $n$ in case of a power--law spectrum
(eq. [37]). Once again the coefficient becomes infinite
when $n\ge -1$.
For $n\approx -2$, $\Cscat\approx 0.3$ so that the scattering around
$\dta_L=0$ is
of the order of
$$\left[\mg V^2(t)\mdL-\mg V(t)\mdL^2\right]^{1/2}\sim 0.3\ V_0\
{\dta_L^2\over\nu^2}.\eqno(47)$$
In such a case the fluctuations around
the spherical collapse are expected to be small.

The fluctuations around the mean density can also be calculated.
They are given by
$$\left[\mg \dta^2(t)\mdL-\mg \dta(t)\mdL^2\right]^{1/2}
\approx \Cscat'\ {\dta_L^2\over\nu^2}\eqno(48)$$
with
$$\eqalign{\Cscat'=
{1\over V_0^2\sigma_L^2}&
\left[\mg \left(V\un(t)\right)^4\md
-2\mg V\deux(t)\left(V\un(t)\right)^2\md+\mg \left(V\deux(t)\right)^2\md\right.
\cr
&\ \ \ -\left.{1\over2}
{\mg \left(V\un(t)\right)^2\,\dta_L^2\md^2-\mg V\deux(t)\,\dta_L^2\md^2
\over \sigma_L^4}\right]^{1/2}.\cr}\eqno(49)$$
For $n\approx-2$ we get $\Cscat'\approx 1$.
The situation obtained there is sketched in Fig.~8:
the overdense objects
are expected to follow the spherical collapse model with a possible departure
and a possible scattering of the same order of magnitude.
Then the accuracy of the spherical collapse model depends on the value
of $\nu$. It is not possible to determine the value of $\dta$ up to which
this approximation is valid since the corrective term is known only
up to the third order in $\dta_L$. At this order the discrepancy between
the dynamics of the spherical collapse and the real dynamics is seen
to be weak as long as $\nu$ is of the order of a few units (say 2, 3)
(eqs. [36, 38]). It implies that the spherical collapse model is relevant
at least up to $\dta\sim 4$ and for values of $\nu$ of a few units.
Whether this approximation breaks down
when $\dta$ is larger than 4 is a problem not solved by these analytic
calculations.

When $n\ge-1$, however, the situation is quite different, and the fluctuations
are unbounded. It seems that the initial overdensity $\dta_L$ does not
determine at all the subsequent evolution of the object which rather
depends on the small--scale fluctuations. The origin of this behavior
is discussed in the next section.

\vskip .5 cm
\leftline{\sl 3.3 Discussion}

I want to stress that
the results presented here take fully into account
the application of a filter function on
the density field at any order of the expansion. It
represents a considerable improvement
compared to previous results. For instance Kofman (1991) proposes
a density distribution function based in the Zel'dovich
(1970) approximation, thought to be valid when the fluctuations
are still small that, however, does not take into account
the filtering of the evolved density field. Bernardeau (1992) also
presents the expected shape of the density distribution
in a quasi--Gaussian approximation (that takes into account
the nonlinear behavior of the density field) with the same
kind of approximation. Both of these calculations were
made for a density in a vanishing volume and appear as approximations
to the exact solution for a finite and large volume. In the calculation
presented here, however, the filtering process has been fully
taken into account. It turns out that the expected nonlinear evolution
of a rare event follows exactly
what would have been expected from the spherical model.

It has been stated that the geometrical
dependences contained in the second or third order terms
may induce a slowing of the dynamics by tidal or shear effects
(Juszkiewicz \& Bouchet 1992) that would produce a sort of previrialization.
The calculations I presented
here take into account these geometrical dependences
but contain no indications of previrialization, at least in the rare
event limit (and as long as $n<-1$). The idea
that the small--scale fluctuations may alter the collapse rate of an object
was expressed by Peebles (1990). He presented numerical experiments
where indeed the growing rate of the fluctuation was slightly lower
compared to the spherical collapse model even for objects of
final density of the
order of 3. This result is at variance with what has been obtained
here in the case of moderately rare events. Previrialization however may be
present in the final stage of the collapse when the radius of the object
starts to decrease for a not extremely rare event.

The extent of the central problem solved in this paper, the expected
behavior of a constrained random field, also deserves some comments.
In the  equation (3) a ``peak'' has been defined as a mere overdense part
of the random density field. The result of part 2, however,
holds even for more constraining initial conditions. One could have
imposed that the volume is at an extremum of the density field (in
case of a rare overdense volume, it is likely to be a maximum).
In such a case the initial random field can be seen as a non--homogeneous
random Gaussian field for which the relation (1) has been replaced by
$$\mg\dta_{\vk_1}\dta_{\vk_2}\md=\dta_3(\vk_1+\vk_2)P(k_1)-
{\vk_1.\vk_2\ P(k_1)P(k_2)\ W(k_1)W(k_2)\over \intk k^2\ W^2(k)\ P(k)}
\eqno(50)$$
where $W(k)$ is the shape of the window function in the Fourier space used
to determine whether a  point is the peak or not (this is not
necessarily the top--hat window function). The random field is still
Gaussian and isotropic around $\vq=0$ so that the whole demonstration
still holds. The corrections to the rare event limit
and in particular the values of the coefficients $\Cexp$ and $\Cscat$, however,
are likely to change (and to be reduced). One could have also required that
a greater volume, $V_l$, centered on $V_0$ contains a certain overdensity,
$\dta_l$ in order to study the formation of objects in a rich, or a poor,
environment as did Bower (1991) in the frame of the Press and Schechter
formalism. The conditional random field is still Gaussian
and isotropic so that the demonstration still holds, and in the rare peak
limit the dynamics
follows the spherical collapse model. The only
change is that the mean value of $\dta_L$ is not zero anymore but
$$\mg \dta_L\md={\int\d^3\vk\ \WTH(k\,R_0)\,\WTH(k\,R_l)\,P(k)\over
\int\d^3\vk\ \WTH^2(k\,R_0)\,P(k)}\ \dta_l\eqno(51)$$
(with an unchanged variance), so that the distribution of $\dta_L$
is shifted towards higher or lower values according to the sign of $\dta_l$.
This is also of great interest
for the Press and Schechter formalism, since as stressed by Bond et al.
(1990) and Blanchard, Valls--Gabaud \& Mamon (1992),
the mass distribution function
is based on an assumption on the dynamics of a somehow isolated
density perturbation: the initial constraint is not only that a given
overdensity is reached in a given volume but also that the surroundings
do not form a higher overdense region. This local constraint may be
especially important for the low values of $\nu$.
This paper does not address, by far, the problem of the exact nonlinear
dynamics with such initial conditions when $\nu$ is small so it
does not provide a justification for the Press and Schechter formula
in its whole generality. The large mass behavior, however, corresponding
to the high $\nu$ limit received a
strong dynamical justification.
The demonstration is unfortunately not complete, since the
virialization process remains beyond the scope of these calculations.
The work of Thomas \& Couchman (1992)
suggests that the spherical collapse model is correct up
to the maximum expansion of the object but fails to reproduce
the contraction rate. This is not that surprising, since in this
regime the
decaying and rotational modes and the shell crossings  that have been
neglected are likely to play a major role in the process of relaxation.

The second major phenomenon highlighted in this paper is the existence of
a critical value for the power--law index. The existence of an unbound
correction when $n\ge-1$ is the signal that a perturbation approach
is not safe. Such a result deserves a deeper examination.
It turns out that the divergence of $\Cexp$ is entirely due
to dynamical properties. For instance if we consider the conditional
expectation values of the spatial derivatives of the displacement field, we
get,
$$\eqalign{
\mg \Psi_{i,j}(\vq)\mdL=&\intk\ {\vk_i\vk_j\over k^2}\
{P(k)\,\WTH(k\,R_0)\over (2\pi)^{3/2}\ \sigma_i^2}\ \dta_L\
e^{\ii\vk.\vq}\cr
&\times
\left[-1+{4\over 21}
\left(\DoDi\right)^2\int{\d^3\vk'\over(2\pi)^{3/2}}P(k')\ f(k'/k)+..\right]\
\cr}\eqno(52)$$
which is finite only if $P(k)\ll k^{-1}$ when $k\to\infty$ (since
$f(\tau)\sim 1/\tau^2$ for $\tau\to\infty$). As a result when $n\ge-1$
the displacement field is not a derivable quantity: two points that
are initially very close may have completely different trajectories.
This is the indication that the matter follows a chaotic behavior rather
than a mere coherent flow towards the overdense regions. It is confirmed by
the derivation of the expectation value of the functional derivative of
the displacement field with the initial fluctuation. The result is
$$\eqalign{
\mg{\delta\ \Psi_{i}(\vq)\over\delta\ \dta_{\vk}}\md
=&\intk\ {-\ii\vk_i\over k^2}\ e^{\ii\vq.\vk}\cr
&\ \left[-\DoDi+{4\over 21}\left(\DoDi\right)^3
 \int{\d^3\vk'\over(2\pi)^{3/2}} P(k')\ f(k'/k)+...\right]\cr}
\eqno(53)$$
which is also divergent for $n\ge-1$. As can be seen in (53) this is
only due to the shape of $f(\tau)$ and not at all to the shape of the
filter function. As a result
a small change in the initial fluctuations would induce a dramatic change
of the behavior of the displacement field. It seems that
the particles that were
initially in the volume $V_0$ do, at least in the first stage of
the dynamics,
a sort of chaotic diffusion, and have
no reason to form ultimately a unique virialized object.
This effect is present whatever the value of $n$ but when
$n<-1$ the correlation length of the displacement field
becomes infinite (as it is for the velocity field) and the displacement
affects the whole object without changing its volume. When $n>-1$
this is no longer the case and small--scale displacements can disrupt
the collapsing peak.
That does not
mean that no object forms at the position of the peak, but it is a
complete nonsense to try to derive its geometrical properties with the
trajectories of the particles that were initially inside $V_0$.
The edge of the volume occupied by the particles is expected to
be extremely fuzzy (like a fractal set), and
the approach adopted here is totally
inadequate to deal with this problem. The use of
more constrained initial conditions (such as in eq. [49]) is not
expected to fix this problem, since these constraints do not
involve the small--scale fluctuations.
A systematic study of the dynamics of constrained fields
with numerical simulations would be helpful to clarify the situation,
but in any case the simple picture used to derive the
mass distribution functions, based on the assumption
of mass conservation, is severely altered in the $n\ge-1$ case.

\vskip .5 cm
\leftline{Acknowledgment:}
The author would like to thank F. Bouchet, R. Juszkiewicz, R. van de Weygaert
and J.R. Bond for useful discussions and the referee, D. H. Weinberg,
for his valuable comments on the original manuscript.

 \vskip 1 cm
 \vfill\eject

\leftline{\bf APPENDIX A: Conditional expectation values
of non-Gaussian variables}
\vskip .5 cm

 This appendix is devoted to a recall of the general form of the
distribution functions of Gaussian and non-Gaussian random variables once
the series of their moments is known.
 We consider two random positive
variables $V_1$ and $V_2$ for which their high--order
correlations and cross--correlations, $\mg V_1^p V_2^q\md$ $p,q$ integers,
are known. These moments define a two variable generating function
$$\mM(\la_1,\la_2)=\sum_{p=0,q=0}^{\infty}
\mg V_1^p V_2^q\md {\la_1^p \la_2^q \over p! q!}.\eqno(A.1)$$
Then the two variable distribution function of $V_1$ and $V_2$ reads
$$\eqalign{
\eta(V_1,V_2)\d V_1\d V_2=\d V_1\d V_2&
\int_{-\ii\infty+0}^{+\ii\infty+0} {\d\la_1\over 2\pi\ii}
\int_{-\ii\infty+0}^{+\ii\infty+0}{\d\la_2\over 2\pi\ii}\cr
&\ \times\mM(\la_1,\la_2)\exp(-\la_1 V_1-\la_2 V_2).\cr}\eqno(A.2)$$

Such a form can be checked by deriving the moments of the distribution
function defined in (A.2). Indeed the expectation
value of $V_1^pV_2^q$ is given by
$$\eqalign{
\int_0^{\infty}\eta(V_1,V_2)\,V_1^p\,V_2^q\,\d V_1\,\d V_2=&
\int_{-\ii\infty+0}^{+\ii\infty+0} {\d\la_1\over 2\pi\ii}{p!\over \la_1^{p+1}}
\int_{-\ii\infty+0}^{+\ii\infty+0}{\d\la_2\over 2\pi\ii}{q!\over \la_2^{q+1}}
\mM(\la_1,\la_2)\cr
=&
\left.{\partial^p\over\partial\la_1^p}{\partial^q\over\partial\la_2^q}
\mM(\la_1,\la_2)\right\vert_{\la_1=0, \la_2=0}
\equiv \mg V_1^pV_2^q\md.\cr}\eqno(A.3)$$

However it
turns out in fact that it is more efficient to work with the cumulants
of the distribution, $\mg V_1^p V_2^q\md_c$.
They are defined recursively from the moments:
$$\eqalign{
\mg 1\md_c&=0;\cr
\mg V\md_c&=\mg V\md;\cr
\mg V^2\md_c&=\mg V^2\md-\mg V\md_c^2;\cr
\mg V_1 V_2\md_c&=\mg V_1 V_2\md-\mg V_1\md_c\ \mg V_2\md_c;\cr
\mg V^3\md_c&=\mg V^3\md-\mg V\md_c^3-3\ \mg V\md_c \mg V^2\md_c;\cr
&\dots\cr}\eqno(A.4)$$
In general the cumulant of order $(p,q)$ is obtained in such a way:
consider all the partitions of a set of $p+q$ points (except the one
constituted by the set itself), take the product of the cumulants
of each of the subsets that come from moments of lower order,
then the cumulant is obtained by substracting
all the terms obtained such a way from the value of the moment of order
$(p,q)$. Note that the usual arithmetic operations are not allowed
on the cumulants and for instance that
$\mg\,A\,B\,C\,\md_c\ne\mg\,(A\,B)\,C\,\md_c$ since the first is the cumulant
of three random variables and the second of two.
The cumulants also define a generating function $\chi(\la_1,\la_2)$
$$\chi(\la_1,\la_2)=\sum_{p=0,q=0}^{\infty}
\mg V_1^p V_2^q\md_c {\la_1^p \la_2^q \over p! q!}.\eqno(A.5)$$
The functions $\chi(\la_1,\la_2)$ and $\mM(\la_1,\la_2)$ are closely
related together by the relation (\eg, Balian \& Schaeffer 1989),
$$\mM(\la_1,\la_2)=\exp\left[\chi(\la_1,\la_2)\right],\eqno(A.6)$$
so that the distribution reads now
$$
\eta(V_1,V_2)\d V_1\d V_2=\d V_1\d V_2
\int_{-\ii\infty+0}^{+\ii\infty+0} {\d\la_1\over 2\pi\ii}
\int_{-\ii\infty+0}^{+\ii\infty+0}{\d\la_2\over 2\pi\ii}
\exp[\chi(\la_1,\la_2)-\la_1 V_1-\la_2 V_2].\eqno(A.7)$$

The general form (A.7) allows one to calculate any statistical quantity
related to the two variables $V_1$ and $V_2$.
One quantity of interest in this paper is the mean value of one
knowing the other, $\mg V_1\md_{V_2}$.
This quantity is given by the ratio,
$$\eqalign{
\mg V_1\md_{V_2}=&\int_0^{\infty}\d V_1\ V_1\ \eta(V_1,V_2)/
\int_0^{\infty}\d V_1\ \eta(V_1,V_2)\cr
=&
\int_{-\ii\infty+0}^{+\ii\infty+0}{\d\la_2\over 2\pi\ii}
\left.{\partial\over\partial \la_1}\chi(\la_1,\la_2)\right\vert_{\la_1=0}
\exp[\chi(0,\la_2)-\la_2 V_2]\cr
&/\int_{-\ii\infty+0}^{+\ii\infty+0}{\d\la_2\over 2\pi\ii}
\exp[\chi(0,\la_2)-\la_2 V_2].\cr}\eqno(A.8)$$
The partial derivative that appears in the result is given by
$$
\left.{\partial\over\partial \la_1}\chi(\la_1,\la_2)\right\vert_{\la_1=0}=
\sum_{p=0}^{\infty}
\mg V_1 V_2^p\md_c {\la_2^p\over p!}.\eqno(A.9)$$

We then apply the relation (A.9) when $V_2$ is a Gaussian variable
of expectation value 1, and of rms fluctuation $\sigma$. This
is the case for the early cosmological density field, so that we
define the random variable $\dta_L\equiv V_2-1$ for such a case.
The function $\chi(0,\la_2)$ then takes a simple form
since
$$\chi_G(0,\la_2)=\la_2+{1\over 2}\sigma^2 \la_2^2\eqno(A.10)$$
(the lowscript $G$ is for Gaussian).
We can also notice that
$$\mg V_1 (1+\dta_L)^p\md_c=\mg V_1 \dta_L^p\md_c.\eqno(A.11)$$
The expectation value of $V_1$ knowing $\dta_L$ is then given by the relation
(A.8) and it reads in this particular case (we make the change of
variable $y=\sigma^2\la_2$),
$$\eqalign{
\mg V_1\md_{\dta_L}=&
\int_{-\ii\infty+0}^{+\ii\infty+0}{\d y\over 2\pi\ii}
\,\mG_{V_1}(y)\,
\exp\left[{(y^2-2\ y\dta_L)\over 2\ \sigma^2}\right]
/\int_{-\ii\infty+0}^{+\ii\infty+0}{\d y\over 2\pi\ii}
\exp\left[{(y^2-2\ y\dta_L)\over(2\ \sigma^2)}\right]\cr}\eqno(A.12)$$
with
$$\mG_{V_1}(y)=\sum_{p=0}^{\infty}
{\mg V_1 \dta_L^p\md_c\over \sigma^{2p}}{y^p\over p!}.\eqno(A.13)$$

We can also compute the expectation value of $V_1^2$ knowing $\dta_L$.
It is given by the following expression,
$$\eqalign{
\mg V_1^2\md_{\dta_L}=&
\int_{-\ii\infty+0}^{+\ii\infty+0}{\d y\over 2\pi\ii}
\left[\mG^2_{V_1}(y)+\mG_{V_1^2}(y)\right]
\exp\left[{(y^2-2\ y\dta_L)\over2\ \sigma^2}\right]\cr
&\ \ \ \ /\int_{-\ii\infty+0}^{+\ii\infty+0}{\d y\over 2\pi\ii}
\exp\left[{(y^2-2\ y\dta_L)\over(2\ \sigma^2)}\right]\cr}\eqno(A.14)$$
with
$$\mG_{V_1^2}(y)=\sum_{p=0}^{\infty}
{\mg V_1^2 \dta_L^p\md_c\over \sigma^{2p}}{y^p\over p!}.\eqno(A.15)$$

The calculation of the integrals (A.12) and (A.14) can be simplified by
the change of variable, $y=\dta_L+\ii z$. As a result
we have
$$\eqalign{\mg V_1\mdL=&
\int_{-\infty}^{+\infty}\d z\,\mG_{V_1}(\dta_L+\ii z)\,
\exp\left[-{z^2\over 2\ \sigma^2}\right]
/\int_{-\infty}^{+\infty}\d z\,
\exp\left[-{z^2\over 2\ \sigma^2}\right]\cr}\eqno(A.16)$$
and
$$\eqalign{\mg V_1^2\mdL=&
\int_{-\infty}^{+\infty}\d z\,
\left[\mG^2_{V_1}(\dta_L+\ii z)+\mG_{V_1^2}(\dta_L+\ii z)\right]
\exp\left[-{z^2\over 2\ \sigma^2}\right]\cr
&\ \ \ \ /\int_{-\infty}^{+\infty}\d z\,
\exp\left[-{z^2\over 2\ \sigma^2}\right].\cr}\eqno(A.17)$$

 \vfill\eject

\leftline{\bf APPENDIX B: Geometrical properties of the top--hat window
function}
\vskip .5 cm

 The purpose is this appendix is to give general properties of the
top--hat window function, $\WTH(\vx)$. It is defined by
$$\eqalign{
\WTH(\vx)&=1 \hbox{  if  } \vert\vx\vert\le R_0;\cr
\WTH(\vx)&=0 \hbox{  otherwise};\cr}\eqno(B.1)$$
for a scale $R_0$.
The Fourier transform of this function is then given by
$$\WTH(k\,R_0)={3\over (kR_0)^3} \left(\sin(kR_0)-kR_0 \cos(kR_0)\right)
\eqno(B.2)$$
where $k$ is the norm of $\vk$.
The latter function will be widely used for the calculation
of the dynamics.

Consider three wave vectors $\vk_1$, $\vk_2$ and $\vk_3$
on which you may have to integrate. The properties of interest
concern the integration over
their angular parts, $\d \Om_1$, $\d \Om_2$ and $\d \Om_3$, and
are the following:
$$\eqalign{
\int\d \Om_1\d\Om_2&\
\WTHd\left[1-\left({\vk_1.\vk_2\over k_1 k_2}\right)^2\right]\cr
&=(4\pi)^2 \ {2\over 3}\ \WTH(k_1\,R_0) \WTH(k_2\,R_0)\cr}\eqno(B.3)$$
and
$$\eqalign{
\int\d \Om_1&\d\Om_2\d\Om_3\
\WTHt\cr
&\times\left[1-\left({\vk_1.\vk_2\over k_1 k_2}\right)^2
-\left({\vk_2.\vk_3\over k_2 k_3}\right)^2
-\left({\vk_3.\vk_1\over k_3 k_1}\right)^2
+2\,{\vk_1.\vk_2\ \vk_2.\vk_3\ \vk_3.\vk_1\over k_1^2 k_2^2 k_3^2 }\right]\cr
&=(4\pi)^3\ {2\over 9}\ \WTH(k_1\,R_0) \WTH(k_2\,R_0) \WTH(k_3\,R_0).\cr}
\eqno(B.4)$$

These properties are essentially some properties of factorizability of the
top--hat window function and are only true for this particular window
function. It means that the top--hat window function commutes with
certain geometrical quantities and that we can replace the angular dependence
of $\vk_1$, $\vk_2$ and $\vk_3$ by their mean as if there was no
window function at all. For instance the mean value of
$1-(\vk_1.\vk_2/k_1 k_2)^2$
in (B.3) is $2/3$ in the absence of window function which exactly corresponds
to the coefficient $2/3$ appearing in (B.3).

The proofs of these properties rely on the fact that $\WTH(k)$
can be written in term of a Bessel function:
$$\WTH(k)=3\sqrt{\pi/2}\ k^{-3/2}\ J_{3/2}(k)\eqno(B.5)$$
(the value of $R_0$ has been set to 1 for simplicity but the demonstrations
obviously hold for any value of $R_0$).
The window function can then be written (\eg, Gradshteyn \& Ryzhik, Eq.
8.532.1),
$$\WTH(\vert\vk_1+\vk_2\vert)=3\pi\sum_{m=0}^{\infty}
({3\over2}+m)(k_1k_2)^{-3/2}\ J_{3/2+m}(k_1)\ J_{3/2+m}(k_2)
{\d\over \d u}P_{m+1}(-u)\eqno(B.6)$$
where $u={\vk_1.\vk_2/ k_1k_2}$ and $P_m$ is a Legendre polynomial.

Then $I_1(k_1,k_2)$ defined as the left part of the relation (B.3) reads
$$\eqalign{I_1(k_1,k_2)
=&3\pi\sum_{m=0}^{\infty}
({3\over2}+m)(k_1k_2)^{-3/2}\ J_{3/2+m}(k_1)\ J_{3/2+m}(k_2)\ (4\pi)\ (2\pi)
\cr
&\ \ \ \ \ \times\int_{-1}^{1}\d u\ (1-u^2)\ {\d\over\d u}P_{m+1}(-u)\cr
=&-3\pi\sum_{m=0}^{\infty}
({3\over2}+m)(k_1k_2)^{-3/2}\ J_{3/2+m}(k_1)\ J_{3/2+m}(k_2)\
(4\pi)^2\cr
&\ \ \ \ \ \times\int_{-1}^{1}\d u\ u\ P_{m+1}(-u).\cr}
\eqno(B.7)$$
The end of the demonstration is straightforward since $u\equiv P_1(u)$,
and the Legendre polynomials are orthogonal to each
other so that $m=0$ is the only non vanishing case. The latter gives
$\int_{-1}^{1}\d u P_1^2(u)=2/3$, so that the integral reads
$$I_1(k_1,k_2)=(4\pi)^2\ 3\pi\ (k_1k_2)^{-3/2}
\ J_{3/2}(k_1)\ J_{3/2}(k_2)\eqno(B.8)$$
which, using the equation (B.5), gives the relation (B.3).

The proof of the relation (B.4) is based on the same kind of calculation
but the demonstration is more complicated due to the fact that three different
vectors are involved. Two steps are required. First of all let me define
$\vK=\vk_1+\vk_2$. The first step will be to separate $\vK$ from
$\vk_3$. I can then define the angles $\Th_{12}$, $\Th_1$, $\Th_2$ and
$\Th_3$ which are respectively the angles between $\vk_1$ and $\vk_2$,
between $\vK$ and $\vk_1$, between $\vK$ and $\vk_2$ and between $\vK$ and
$\vk_3$ (note that $\Th_{12}=\Th_1+\Th_2$).
The vector $\vk_3$ is also defined by its recession angle, $\Phi_3$,
around the vector $\vK$ (they are set to 0 for the vectors $\vk_1$ and
$\vk_2$). The angular part appearing in the left part of (B.4) is
$$\eqalign{
A(\Th_1,\Th_2,\Th_3,\Phi_3)
=1&-\left({\vk_1.\vk_2\over k_1 k_2}\right)^2
-\left({\vk_2.\vk_3\over k_2 k_3}\right)^2
-\left({\vk_3.\vk_1\over k_3 k_1}\right)^2\cr
&+2\,{\vk_1.\vk_2\ \vk_2.\vk_3\ \vk_3.\vk_1\over k_1^2 k_2^2 k_3^2 }\cr
=1&-\cos^2(\Th_{12})\cr
&-\left[\cos(\Th_3)\cos(\Th_1)+\sin(\Th_3)\sin(\Th_1)\sin(\Phi_3)\right]^2\cr
&-\left[\cos(\Th_3)\cos(\Th_2)+\sin(\Th_3)\sin(\Th_2)\sin(\Phi_3)\right]^2\cr
&+2\cos(\Th_{12})
\left[\cos(\Th_3)\cos(\Th_1)+\sin(\Th_3)\sin(\Th_1)\sin(\Phi_3)\right]\cr
&\ \ \ \ \ \times
\left[\cos(\Th_3)\cos(\Th_2)+\sin(\Th_3)\sin(\Th_2)\sin(\Phi_3)\right].\cr}
\eqno(B.9)$$
The integration over $\Phi_3$ can be made at this stage since
the factor ${\WTH(\vert\vk_1+\vk_2+\vk_3\vert)}$ is independent of $\Phi_3$.
The result reads
$$\eqalign{
B&(\Th_1,\Th_2,\Th_3)=\int_{0}^{2\pi}\d \Phi_3\ A(\Th_1,\Th_2,\Th_3,\Phi_3)=\cr
&2\pi\ \left(
1-\cos^2(\Th_{12})
-\cos^2(\Th_3)\cos^2(\Th_1)-{1\over2}\sin^2(\Th_3)\sin^2(\Th_1)\right.\cr
&\ \ \ \ \ \
\left.-\cos^2(\Th_3)\cos^2(\Th_2)-{1\over2}\sin^2(\Th_3)\sin^2(\Th_2)\right.\cr
&\ \ \ \ \ \
\left.+2\cos(\Th_{12})\left[\cos^2(\Th_3)\cos(\Th_1)\cos(\Th_2)+{1\over2}
\sin^2(\Th_3)\sin(\Th_1)\sin(\Th_2)\right]\right).\cr}\eqno(B.10)$$
The function $I_2(k_1,k_2,k_3)$, defined as the left part of (B.4), then
reads
$$\eqalign{
I_2(k_1,k_2,k_3)=\int&\d\Om_1\ \d\Om_2\ \sin(\Th_3)\d\Th_3
\cr
&\times 3\pi\ \sum_{m=0}^{\infty}
({3\over2}+m)(Kk_3)^{-3/2}\ J_{3/2+m}(K)\ J_{3/2+m}(k_3)\cr
&\times
{\d\over \d u}P_{m+1}(-\cos(\Th_3))\ B(\Th_1,\Th_2,\Th_3).\cr}\eqno(B.11)$$
The integration over $\Th_3$ takes advantage of the fact that
$B(\Th_1,\Th_2,\Th_3)$ contains only terms independent of $\Th_3$ or
proportional to $\cos^2(\Th_3)$. We can then use general
properties of the Legendre polynomials,
$$\eqalign{
&\int_{-1}^{1}\d u{\d\over\d u}P_{m+1}(-u)=2\hbox{  for even values of }m;\cr
&\int_{-1}^{1}\d u{\d\over\d u}P_{m+1}(-u)\ u^2=2\hbox{  for even values
of }m\hbox{ and }m\ne 0;\cr
&\int_{-1}^{1}\d u{\d\over\d u}P_{1}(-u)\ u^2=2/3;\cr}\eqno(B.12)$$
and these integrals are zero otherwise,
for $u=\cos(\Th_3)$ to compute (B.11). When $m\ne0$, the two
contributions coming from the terms independent of $\Th_3$ and the
ones proportional to $\cos^2(\Th_3)$ have to be added which is obtained by
setting $\Th_3=0$ in $B(\Th_1,\Th_2,\Th_3)$ which gives zero.

For $m=0$ the contribution of the integral has to be calculated directly.
Taking advantage of the fact that $\Th_{12}=\Th_1+\Th_2$ we obtain
$$\int \sin(\Th_3)\d \Th_3\ {\d\over \d u} P_{1}(-\cos(\Th_3))
\ B(\Th_1,\Th_2,\Th_3)={4\pi\over 3}\ (1-\cos^2(\Th_{12}))$$
so that
$$\eqalign{
I_2(k_1,k_2,k_3)&=\int\d\Om_1\ \d\Om_2\
3\pi
{3\over2}(Kk_3)^{-3/2}\ J_{3/2}(K)\ J_{3/2}(k_3)
{4\pi\over 3} \left[1-\cos^2(\Th_{12})\right]\cr
&={4\pi\over3}\ \WTH(k_3)\ I_1(k_1,k_2)\cr}
\eqno(B.13)$$
which using the equation (B.3) gives straightforwardly the relation (B.4).

 \vfill\eject

\leftline{\bf APPENDIX C: Calculation of $\mG_V(t,y)$, $\mG_{V^2}(t,y)$
and $\mG_{\dta}(t,y)$}

The function $\mG_V(t,y)$ is defined in equation (23) and will be calculated
for $\sigma_L=0$ that is with the approximation (26) for the series
of the coefficient $j_p$. Let me first sketch the principle of this
calculation. The starting point is the dynamical
equations in Lagrangian space describing the evolution of a pressureless fluid
(eqs. [C.2,C.3]). Then they will be written in Fourier space displaying
the geometrical dependences examined in the previous appendix. Taking the
cumulants of each term (once multiplied by $\dta_L$ at the required power)
introduces integrations over the angles of the wave vectors that are
then known from the result of the appendix B.
As the geometrical dependences vanish we end up with a system
of equations of which $\mG(t,y)$ is solution. The field equation system
have thus been replaced by a simple first order differential system.

\vskip .5 cm

\leftline{\sl C.1 Notations and general properties}

The fulfillment of this derivation requires the definitions of new
quantities presented in the following. Let me however start with a general
property of the cumulants widely used
throughout this appendix.
In this paragraph any quantity denoted $Q^{(p)}$ stands for the
order $p$ of $Q$ relatively to the random variables $\dta_{\vk}$.
The property of interest concerns the cumulant
of a product. If $Q$ can be
written as a product, $Q=F_1\ F_2$, the order $p$ of $Q$ is given by
${Q^{(p)}=\sum_{r=0}^p\,F_1^{(p-r)}\ F_2^{(r)}}$ and the cumulant
it gives when multiplied by $\dta_L^p$ is
$$\mg Q^{(p)}\dta_L^p\md_c=\sum_{r=0}^p{p!\over r!(p-r)!}
\mg F_1^{(p-r)} \dta_L^{p-r}\md_c\ \mg F_2^{(r)} \dta_L^{r}\md_c.\eqno(C.1)$$
This result is due to the properties of the Gaussian variables. A product
of $2p$ Gaussian variables has an expectation value that is written as
a summation of $p$ moments of pairs, so that the terms involved
in (C.1) separate in two sets: the factors involving $F_1$ and the factors
involving $F_2$. The symmetry factor $p!/r!/(p-r)!$ is due to the
number of possibilities you have to separate a set of $p$ points
into two sets of $r$ and $p-r$ points.

The starting point of the derivation of $\mG_V(t,y)$ is the motion equations
derived in the main section, eqs. (17, 20), that I recall here
$$\eqalign{
J(t,\vq)=&1+\gradq.\Psi+{1\over 2}\left[\left(\gradq.\Psi\right)^2-
\sum_{ij}\Psi_{i,j}\Psi_{j,i}\right]\cr
&+{1\over 6}
\left[\left(\gradq.\Psi\right)^3-3\gradq.\Psi\sum_{ij}\Psi_{i,j}\Psi_{j,i}
+2 \sum_{ijk}\Psi_{i,j}\Psi_{j,k}\Psi_{k,i}\right]\cr}\eqno(C.2)$$
where $\gradq$ is the gradient taken relatively to $\vq$.
The second equation is
$$\eqalign{
\Der\left(\gradq.\Psi\right)+
\gradq.\Psi\,\Der\left(\gradq.\Psi\right)-&
\sum_{ij}\Psi_{i,j}\Der\left(\Psi_{j,i}\right)
+{1\over 2}\left(\gradq.\Psi\right)^2\Der\left(\gradq.\Psi\right)\cr
-{1\over2}\Der\left(\gradq.\Psi\right)\sum_{ij}\left[\Psi_{i,j}\Psi_{j,i}
\right.
-&\left.\gradq.\Psi\,\Der\left(\Psi_{i,j}\right)\Psi_{j,i}\right]
+\sum_{ijk}\Psi_{i,j}\Psi_{j,k}\Der\left(\Psi_{k,i}\right)\cr
&\ \ \ =-4\pi G\rhob\left[1-J(t,\vq)\right]\cr}
\eqno(C.3)$$
where $\Der(.)$ is a time derivative operator,
$$\Der(A)=\ddot{A}+2{\dot{a}\over a}\dot{A}.\eqno(C.4)$$
The displacement field has also to be irrotational in the
real space.
These equations completely define the time evolution of the
displacement field. A calculation of the first orders when the displacement
field is considered as a small quantity is given in the appendix D.
But such direct perturbative calculations are inadequate to compute
the values of $j_p(t)$ as defined in (24). The series of the parameters
will be obtained as a functional equation for $\mG_V(t,y)$.

 For convenience I define few other quantities. Similarly to $\mG_V$
I can consider the function
$$\mG_D(t,y)=\sum_{p=0}^{\infty} {\mg \left[1/V_0\int\d^3\vq\, \gradq.
\Psi(t,\vq)
\right] \dta_L^p\md_c\over \sigma_L^{2p}}{y^p\over p!}\eqno(C.5)$$
and the parameters
$$d_p(t)\equiv\mg \left({1\over V_0}\int\d^3\vq\, \gradq.\Psi(t,\vq)\right)
 \dta_L^p\md_c=\mg \left({1\over V_0}\int\d^3\vq\, \gradq.\Psi^{(p)}(t)\right)
 \dta_L^p\md_c,\eqno(C.6)$$
in the rare peak approximation.
I also denote $\jk^{(p)}(t)$ and $\dk^{(p)}(t)$ the Fourier transform
of respectively $J(t,\vq)$ and $\gradq.\Psi(t,\vq)$ at order
$p$ in an expansion with respect to the random variables $\dta_{\vk}$.
All these
quantities are time dependent and the operator $\Der(.)$ defined in (C.3)
can apply to any of them.

The term of order $p$ for the volume then reads
$$V^{(p)}(t)/V_0=\int{\d^3\vk\over (2\pi)^{3/2}}
\, \WTH(k\,R_0)\ \jk^{(p)}(t)\eqno(C.7)$$
and we have
$$\eqalign{
j_p=&\int{\d^3\vk\over (2\pi)^{3/2}}
\mg\jk^{(p)}(t)\,\dta_L^p\md_c \WTH(k\,R_0).\cr}\eqno(C.8)$$
We have similarly for the divergence of the displacement field,
$${1\over V_0}\int_{V_0}\d^3\vq\, \gradq.\Psi^{(p)}(t,\vq)=
\int{\d^3\vk\over (2\pi)^{3/2}}\, \WTH(k\,R_0)\ \dk^{(p)}(t).\eqno(C.9)$$
and
$$\eqalign{
d_p=&\int{\d^3\vk\over (2\pi)^{3/2}}
\mg\dk^{(p)}(t)\,\dta_L^p\md_c \WTH(k\,R_0).\cr}\eqno(C.10)$$
The components of the matrix $\Psi_{i,j}$ also define some cumulants,
$$s_{p;i,j}(t,\vq)=\mg \Psi^{(p)}_{i,j}(t,\vq) \dta_L^p\md_c.\eqno(C.11)$$
We can notice that the parameters $j_p(t)$ can simply be expressed with
the quantities $s_{p;i,j}$ since
$$j_p(t)={1\over V_0}\int_{V_0}\d^3\vq \sum_{i}s_{p;i,i}(t,\vq).\eqno(C.12)$$

Written in $\vk$-space the equations (C.2) and (C.3) introduce the functions
$$J_1(\vk_1,\vk_2)=
\WTHd
\left[1-\left({\vk_1.\vk_2\over k_1 k_2}\right)^2\right],\eqno(C.13)
$$
and
$$\eqalign{J_2&(\vk_1,\vk_2,\vk_3)=
\WTHt\cr
&\ \ \ \ \ \times\left[1-\left({\vk_1.\vk_2\over k_1 k_2}\right)^2
-\left({\vk_2.\vk_3\over k_2 k_3}\right)^2
-\left({\vk_3.\vk_1\over k_3 k_1}\right)^2
+2\,{\vk_1.\vk_2\ \vk_2.\vk_3\ \vk_3.\vk_1\over k_1^2 k_2^2 k_3^2 }\right].\cr}
\eqno(C.14)$$
These functions have been studied in the appendix B where it is shown that
$$\int\d\Omega_1\ \d\Omega_2\ J_1(\vk_1,\vk_2)=(4\pi)^2\ {2\over3}
\WTH(k_1\,R_0)\ \WTH(k_2\,R_0)\eqno(C.15)$$
and
$$\int\d\Omega_1\ \d\Omega_2\ \d\Omega_3
\ J_2(\vk_1,\vk_2,\vk_3)=(4\pi)^3\ {2\over9}
\WTH(k_1\,R_0)\ \WTH(k_2\,R_0)\ \WTH(k_3\,R_0).\eqno(C.16)$$

\vskip .5 cm
\leftline{\sl C.2 The function $\mG_V(t,y)$}

The first step of the demonstration is to remark that the quantities
$s_{p;i,j}(t,\vq)$ are symmetric in $i,j$ and are functions of $\vert\vq\vert$
only. The fact that the initial conditions are randomly isotropic
around $\vq=0$ implies directly that $s_{p;i,j}$
is a  function of the norm of $\vq$. Moreover when $i\ne j$,
$s_{p;i,j}-s_{p;j,i}$
are the components of a vector (given by the rotational of the displacement
field) so that
these quantities should be zero for the same reason:
no particular direction exists in the initial conditions. Note that
this property
is correct at the level of these expectation values but not
at all for the random displacement field itself. The random variable
$\Psi_{i,j}-\Psi_{j,i}$ is not zero although some of its expectation
values vanish. The implication of such property is that $s_{p;i,j}$
can be written
$$s_{p;i,j}(t,\vq)=\int{\d^3\vk\over (2\pi)^{3/2}}
{\vk_i\vk_j\over k^2}s_p(k)\ e^{\ii \vk.\vq}\eqno(C.17)$$
and, according to (C.12), $s(k)$ is then simply given by
$$s_p(k)=
\mg\dk^{(p)}(t)\,\dta_L^p\md_c\eqno(C.18)$$

We can now transform the equations (C.2) and (C.3) involving
the displacement field in two equations between the series $j_p$ and $d_p$.

Let me start with the equation (C.2). This equation has to be true at
any order so that for $p\ne0$,
$$\eqalign{
J^{(p)}(t,\vq)=&\gradq.\Psi^{(p)}+{1\over 2}\sum_{r=0}^p
\left[\gradq.\Psi^{(p-r)}\gradq.\Psi^{(r)}-
\sum_{ij}\Psi_{i,j}^{(p-r)}\Psi^{(r)}_{j,i}\right]\cr
&+{1\over 6}\sum_{r=0,s=0}^{r=p,s=p-r}
\bigg[\gradq.\Psi^{(p-r-s)}\gradq.\Psi^{(r)}\gradq.\Psi^{(s)}\cr
&-3\gradq.\Psi^{(p-r-s)}\sum_{ij}\Psi_{i,j}^{(r)}\Psi_{j,i}^{(s)}
+2 \sum_{ijk}\Psi_{i,j}^{(p-r-s)}\Psi_{j,k}^{(r)}\Psi_{k,i}^{(s)}\bigg]\cr
}\eqno(C.19)$$

The equation (C.19) can now be transformed in an equation between the cumulants
$j_p$ and $s_{p;i,j}$ by multiplying it by $\dta_L^p$ and taking
the cumulant of each term. Using the equation (C.1), (C.17) and (C.18)
we obtain
$$\eqalign{
&\int{\d^3\vk\over(2\pi)^{3/2}}\mg j_{\vk}^{(p)}\dta_L^p\md_c\ e^{\ii\vk.\vq}
=\int{\d^3\vk\over(2\pi)^{3/2}}s_p(k)\ e^{\ii\vk.\vq}\cr
&+{1\over 2}\sum_{r=0}^p {p!\over r!(p-r)!}\int{\d^3\vk_1\over(2\pi)^{3/2}}
{\d^3\vk_2\over(2\pi)^{3/2}}
s_{p-r}(k_1)\,s_{r}(k_2)\left[1-\left({\vk_1.\vk_2\over k_1k_2}\right)^2\right]
\ e^{\ii(\vk_1+\vk_2).\vq}\cr
&+{1\over 6}\sum_{r=0,s=0}^{r=p,s=p-r}{p!\over (p-r-s)!r!s!}\intt
s_{p-r-s}(k_1)s_{r}(k_2)s_{s}(k_3)\cr
&\times\left[1-\left({\vk_1.\vk_2\over k_1 k_2}\right)^2
-\left({\vk_2.\vk_3\over k_2 k_3}\right)^2
-\left({\vk_3.\vk_1\over k_3 k_1}\right)^2
+2\,{\vk_1.\vk_2\ \vk_2.\vk_3\ \vk_3.\vk_1\over k_1^2 k_2^2 k_3^2 }\right]
\cr
&\times e^{\ii(\vk_1+\vk_2+\vk_3).\vq}.\cr}\eqno(C.20)$$
The average of (C.20) over the volume $V_0$ introduces the
parameters $j_p$, $d_p$ and the functions $J_1$ and $J_2$ defined in
(C.13) and (C.14):
$$\eqalign{
&j_p=d_p
+{1\over 2}\sum_{r=0}^p {p!\over r!(p-r)!}\int{\d^3\vk_1\over(2\pi)^{3/2}}
{\d^3\vk_2\over(2\pi)^{3/2}}s_{p-r}(k_1)\,s_{r}(k_2)J_1(\vk_1,\vk_2)\cr
&+{1\over 6}\sum_{r=0,s=0}^{r=p,s=p-r}{p!\over (p-r-s)!r!s!}\intt
s_{p-r-s}(k_1)s_{r}(k_2)s_{s}(k_3)\cr
&\times J_2(\vk_1,\vk_2,\vk_3).\cr}\eqno(C.21)$$
which can be integrated out by the use of the properties
(C.15) and (C.16). We eventually get
$$j_p=d_p+{1\over3}\sum_{r=0}^{r=p}{p!\over (p-r)!r!}d_r\,d_{p-r}
+{1\over 27}\sum_{r=0,s=0}^{r=p,s=p-r}{p!\over r!s!(p-r-s)!}
d_r\ d_s\ d_{p-r-s}.\eqno(C.22)$$
Then, using the definitions (20) and (C.4)
we can derive a relation between $\mG_V$ and $\mG_D$,
$$\mG_V(t,y)/V_0=
1+\mG_D(t,y)+{1\over 3}\mG_D^2(t,y)+{1\over 27}\mG_D^3(t,y)=
\left[1+\mG_D(t,y)/3\right]^3.\eqno(C.23)$$

 The equation (C.3) leads to another relationship of the same
kind.
The only qualitative change is the presence of the operator
$\Der(.)$. But as it is a linear operator, it commutes with any
operations such as the Fourier transform, ensemble average,
and integration over $V_0$. Similar calculations then lead to
$$\eqalign{
\Der(d_p)+&{2\over3}\sum_{r=0}^{r=p}{p!\over (p-r)!r!}d_r\,\Der(d_{p-r})+
{1\over 9}\sum_{r=0,s=0}^{r=p,s=p-r}{p!\over r!s!(p-r-s)!}
\Der(d_r)\,d_s\,d_{p-r-s}\cr
&=4\pi\ G\ \rhob\ j_p\cr}\eqno(C.24)$$
which gives
$$\Der\left[\mG_D(t,y)\right]
\left[1+\mG_D(t,y)/3\right]^2
=-4\pi\ G\ \rhob\left[V_0-\mG_V(t,y)\right].\eqno(C.25)$$
The equations (C.23) and (C.25) with the definition of $\Der(.)$
lead to the differential equation
$${\d^2\over\d t^2}\left(R_0 \left[\mG_V(t,y)/V_0\right]^{1/3}\right)=
{-(G\ \rhob+\Lambda/4\pi)\ V_0\over R_0^2 \left[\mG_V(t,y)/V_0\right]^{2/3}}
\eqno(C.26)$$
since the comoving radius $R_0(t)$ is proportional to the expansion factor
and satisfies
 $\ddot{R_0}=-4\pi/3\ \rhob\,G\,R_0+\Lambda R_0/3$. The equation (C.26)
is the differential equation describing a spherical collapse,
where ${R_0 \left[\mG_V(t,y)/V_0\right]^{1/3}}$ is the size
of the object. The behavior of $\mG_V(y)$ for small values of $y$
ensures that the corresponding overdensity is simply $y$.
The function $\mG_V(t,y)$ is then the volume occupied at the time $t$
by an object of linear initial overdensity $y$ and of
initial comoving volume $V_0$.

This result is not so surprising since we saw that the application of
the top--hat window function just replaces the geometrical dependences
that enter in the real dynamics by their means, \ie, their monopole
part, and the dynamics of the spherical collapse just corresponds
to the case where there is only one monopole.

\vskip .5 cm
\leftline{\sl C.3 The functions $\mG_{V^2}(t,y)$ and $\mG_{\dta}(t,y)$}

 The purpose of this part is to compute the values of the functions,
$$\eqalign{
\mG_{V^2}(t,y)&=\sum_{p=0}^{\infty}
{\mg V^2(t) \dta_L^p\md_c\over \sigma_L^{2p}}{y^p\over p!},\cr
\mG_{\dta}(t,y)&=\sum_{p=0}^{\infty}
{\mg [V_0/V(t)-1] \dta_L^p\md_c\over \sigma_L^{2p}}{y^p\over p!},\cr}
\eqno(C.27)$$
in the rare peak approximation. The first function involves a cumulant
between $p+2$ variables that requires
at least $p+1$ links between them. It can only be
done with the moment of $2p+2$ Gaussian variables. As a result
$$\mg V^2(t) \dta_L^p\md_c\approx
\sum_{r=1}^{r=p+1}\mg V^{(r)}(t)V^{(p+2-r)}(t)
\dta_L^p\md_c.\eqno(C.28)$$
The calculation of such quantities is simple but requires a good understanding
of the result obtained in the previous part.
The order $p$ of the volume can be written in such a way,
$$\eqalign{
V^{(p)}(t)=V_0
\int{\d^3\vk_1\over(2\pi)^{3/2}}\dots{\d^3\vk_p\over(2\pi)^{3/2}}
&\dta_{\vk_1}\dots\dta_{\vk_p}
\WTH\left[\vert\vk_1+\dots+\vk_p\vert\ R_0\right]\cr
&\times J(\vk_1,\dots,\vk_p)\cr}\eqno(C.29)$$
where $J(\vk_1,\dots,\vk_p)$ is a function of the angular parts
of the wave vectors
$\vk_1,\dots,\vk_p$ only.
The result obtained in the part (C.1) is then a property of the
function $J(\vk_1,\dots,\vk_p)$ that reads
$$\eqalign{
\int\d\Omega_1\dots\d\Omega_p&\WTH\left[\vert\vk_1+\dots+\vk_p\vert\ R_0\right]
J(\vk_1,\dots,\vk_p)=\cr
&(4\pi)^p\ {j_p(t)\over \sigma_L^p}\
\WTH(k_1\,R_0)\dots \WTH(k_p\,R_0).\cr}\eqno(C.30)$$
Indeed the function $J(\vk_1,\dots,\vk_p)$ turns out to be a combination of the
functions $J_1$ and $J_2$ for which such a factorization has been explicitly
established in the appendix B.
The relation (C.30) can also be used in (C.28) since the angular parts
of the wave vectors appearing in the expressions of $V^{(r)}$
and of $V^{(p+2-r)}$ are independent. As a result we simply obtain
$$\mg V^2(t)\ \dta_L^p\md_c=
\sum_{r=1}^{r=p+1}{p!\over (r-1)!(p+1-r)!}
j_r(t)\,j_{p+2-r}(t)\eqno(C.31)$$
so that
$$\mG_{V^2}(t,y)=\sigma_L^2\left[{\d\over\d y}
\mG_V(t,y)\right]^2.\eqno(C.32)$$
The same geometrical arguments can be invoked for the calculation
of $\mG_{1/V}(t,y)$ that lead to the expression,
$$\mG_{\dta}(t,y)=V_0/\mG_V(t,y)-1.\eqno(C.33)$$

 \vfill\eject

\leftline{\bf APPENDIX D: Perturbative calculations in Lagrangian
coordinates}

This Appendix is devoted to the derivation of the third order of the
Jacobian in a Lagrangian calculation (see the main text for details), and
some expectation values of interest. The principle of these calculations
is not new and has been presented elsewhere (see for instance Moutarde \etal
1991, Buchert 1992)

\vskip .5 cm
\leftline{\sl D.1 The three first orders of the displacement field}

To each comoving position $\vq$ is associated a displacement vector
$\Psi(\vq)$ giving the present position $\vx$ of the particle being at
$\vq$ initially,
$$\vx=\vq+\Psi(\vq),\eqno(D.1)$$
and the displacement field is thought, in a perturbative approach, to be
a small correction to the Hubble expansion (contained in $\vq$).

The Jacobian of the transformation between $\vq$ and
$\vx$, $\vert \partial \vx/\partial \vq\vert$, gives
a direct information on the
volume occupied by the particles at any stage of the dynamics
(before shell crossing).
Its expression is obtained by the calculation of the
determinant of the matrix the elements of which are

$${\partial \vx_i\over \partial \vq_j}=\dta_{ij}+{\partial \Psi_i \over
\partial q_j}\eqno(D.2)$$
where $\dta_{ij}$ is the Kronecker symbol.
It leads to the relation,
$$\eqalign{
J(t,\vq)=&1+\gradq.\Psi+{1\over 2}\left[\left(\gradq.\Psi\right)^2-
\sum_{ij}\Psi_{i,j}\Psi_{j,i}\right]\cr
&+{1\over 6}
\left[\left(\gradq.\Psi\right)^3-3\gradq.\Psi\sum_{ij}\Psi_{i,j}\Psi_{j,i}
+2 \sum_{ijk}\Psi_{i,j}\Psi_{j,k}\Psi_{k,i}\right],\cr}\eqno(D.3)$$
where $\gradq$ is the gradient taken relatively to $\vq$
and the summations are made over the three spatial components.
The divergence of the acceleration field is proportional to the
density field so that
$$J(t,\vq)\left(\gradx.\ddot{\vx}+2{\dot{a}\over a}\gradx.\dot{\vx}\right)
=-4\pi G
\rhob\left[1-J(t,\vq)\right]\eqno(D.4)$$
in which a dot denotes a time derivative. In term of the displacement field
it reads
$$\eqalign{
\Der\left(\gradq.\Psi\right)+
\gradq.\Psi\,\Der\left(\gradq.\Psi\right)-&
\sum_{ij}\Psi_{i,j}\Der\left(\Psi_{j,i}\right)
+{1\over 2}\left(\gradq.\Psi\right)^2\Der\left(\gradq.\Psi\right)\cr
-{1\over2}\Der\left(\gradq.\Psi\right)\sum_{ij}\left[\Psi_{i,j}
\Psi_{j,i}\right.
-&\left.\gradq.\Psi\,\Der\left(\Psi_{i,j}\right)\Psi_{j,i}\right]
+\sum_{ijk}\Psi_{i,j}\Psi_{j,k}\Der\left(\Psi_{k,i}\right)\cr
&\ \ \ =-4\pi G\rhob\left[1-J(t,\vq)\right]\cr}
\eqno(D.5)$$
where $\Der(.)$ is a time derivative operator,
$$\Der(A)=\ddot{A}+2{\dot{a}\over a}\dot{A}.\eqno(D.6)$$
The displacement field is completely determined by the
the non--rotational constraint in the $\vx$ coordinates, which leads to a
third equation,
$$\eqalign{
&\dot{\Psi}_{i,j}-\dot{\Psi}_{j,i}+\gradq.\Psi\
(\dot{\Psi}_{i,j}-\dot{\Psi}_{j,i})-
\sum_{k}\dot{\Psi}_{i,k}\Psi_{k,j}+\sum_{k}\dot{\Psi}_{j,k}\Psi_{k,i}\cr
&+{1\over2}\left[(\gradq.\Psi)^2-\sum_{ij}
\Psi_{i,j}\Psi_{j,i}\right](\dot{\Psi}_{i,j}-\dot{\Psi}_{j,i})
-\gradq.\Psi
\sum_{k}\left(
\dot{\Psi}_{i,k}\Psi_{k,j}-\dot{\Psi}_{j,k}\Psi_{k,i}
\right)\cr
&+\sum_{kl} \left(
\dot{\Psi}_{i,k}\Psi_{k,l}\Psi_{l,j}-\dot{\Psi}_{j,k}\Psi_{k,l}\Psi_{l,i}
\right)=0.\cr}\eqno(D.7)$$

{\sl In the following I will assume that the
cosmological parameters are the ones of an Einstein-de Sitter universe.}

The principle of the calculation is the following. We assume
that the displacement field is a small quantity. At a given order the
elimination of $J(t,\vq)$ in (D.3) and (D.5) provides an equation
for the divergence of the displacement field. The equation (D.7) then
gives the complete information over its non--symmetric part.

The time dependence of the first order solution, $D(t)$, is solution
of the linearized equations in the displacement field.
As a result $D(t)$ is solution of
$$\Der\left(D(t)\right)=4\pi\,G\,\rhob\,D(t).\eqno(D.8)$$
There are two solutions for the previous equations: One is an increasing
function of time and is proportional to $a(t)$ the other
is a decreasing function of time, proportional to $t^{-1}$.
Since we assume that
the fluctuations begin to grow at an arbitrarily small time the growing
mode completely dominates the dynamics.

The first order solution then reads
$$J\un(t,\vq)=\gradq.\Psi\un(t,\vq)=D(t)\epsilon_+(\vq),\eqno(D.9)$$
where $\epsilon_+(\vq)$ is a small perturbation in the displacement field
corresponding to the growing mode.
This quantity is related to the initial overdensity that gives birth to the
cosmic structures. The equation (D.7) insures that
$\Psi_{i,j}\un=\Psi_{j,i}\un$ so that the displacement field is
completely determined by its divergence.

At the second order the equation for the divergence is
$$\eqalign{
&\Der\left(\gradq.\Psi\deux\right)+\gradq.\Psi\un(\vq,t)\Der\left(
\gradq.\Psi\un\right)-\Psi_{i,j}\un.\Der\left(\Psi_{i,j}\un\right)=\cr
&\ \ \ 4 \pi G \rhob\left[\left(\gradq.\Psi\un\right)^2
-\sum_{ij}\left(\Psi_{i,j}\un\right)^2\right].\cr}\eqno(D.10)$$
The time dependence of the solution of this equation is $D^2(t)$
(once the only pure growing mode has been selected). Note that this result
is specific to the Einstein-de Sitter universe and is not true in
general. The exact result for any value of $\Omega$ is given by
Bouchet \etal (1992).
As a result we obtained,
$$\eqalign{
\gradq.\Psi\deux(t,\vq)&={-3\over 14}
\left[\left(\gradq.\Psi\un\right)^2
-\sum_{ij}\left(\Psi_{i,j}\un\right)^2\right],\cr
J\deux(t, \vq)&={2 \over 7}
\left[\left(\gradq.\Psi\un\right)^2
-\sum_{ij}\left(\Psi_{i,j}\un\right)^2\right].\cr}\eqno(D.11)$$
Once again the equation (D.7) insures that $\Psi\deux_{i,j}=\Psi\deux_{j,i}$.

The third order is calculated the same way. Its time dependence
is simply $D(t)^3$ and the solution
reads
$$\eqalign{
\gradq.\Psi\trois(t,\vq)=&{-5 \over 9}
\left[\gradq.\Psi\un \gradq.\Psi\deux
-\sum_{ij}\Psi_{i,j}\un\Psi_{i,j}\deux\right]\cr
&{-1\over 18}
\left[\left(\gradq.\Psi\un\right)^3-3 \gradq.\Psi\un\sum_{ij}
\left(\Psi_{i,j}\un\right)^2+\sum_{ijk}\Psi_{i,j}\un
\Psi_{j,k}\un\Psi_{k,i}\un\right],\cr
J\trois(t,\vq)=&{4\over 9}
\left[\gradq.\Psi\un \gradq.\Psi\deux
-\sum_{ij}\Psi_{i,j}\un\Psi_{i,j}\deux\right]\cr
&+{1\over 9}
\left[\left(\gradq.\Psi\un\right)^3-3 \gradq.\Psi\un\sum_{ij}
\left(\Psi_{i,j}\un\right)^2+\sum_{ijk}\Psi_{i,j}\un
\Psi_{j,k}\un\Psi_{k,i}\un\right].\cr}\eqno(D.12)$$
Note that $\Psi\trois_{i,j}\ne\Psi\trois_{j,i}$ so that a complete
determination of $\Psi\trois_{i,j}$ cannot be derived from (D.12).
\vskip .5 cm
\leftline{\sl D.2 Calculation of $\mg V^{(3)}(t)\dta_L\md$}

The second part of this appendix is devoted to the calculation of
$\mg V\trois(t)\,\dta_L\md$, where $V\trois(t)$
is the integral of $J\trois$ over the volume $V_0$.
In the wave vectors space these quantities are given by integrations
over $\vk$ with the filter function $\WTH(k\ R_0)$ as defined
in the appendix B.
 The Fourier transforms, $\jk\un$,
of the first order of the Jacobian are defined by
$$J\un(\vq)=\int{\d^3 \vk\over (2\pi)^{3/2}}\ \jk\un\ \exp(\ii \vq.\vk).
\eqno(D.13)$$
As a result the integral of $J\un$ over the volume is given by
$$V\un=V_0\int {\d^3 \vk \over (2 \pi)^{3/2}}
\jk\un\ \WTH(k\ R_0) \eqno(D.14)$$
and
$$\eqalign{
V\trois=V_0\int&
{\d^3 \vk_1 \over (2 \pi)^{3/2}}{\d^3 \vk_2 \over (2 \pi)^{3/2}}
{\d^3 \vk_3 \over (2 \pi)^{3/2}}
\WTH\left(\vert\vk_1+\vk_2+\vk_3\vert\ R_0\right)
\cr
&\times j_{\vk_1}\un j_{\vk_2}\un j_{\vk_3}\un \left[
{-2 \over 21}\left(1-
{\left(\vk_1.(\vk_2+\vk_3)\right)^2 \over k_1^2 \vert\vk_2+\vk_3\vert^2}\right)
\left(1-\left[{\vk_2.\vk_3\over k_2 k_3}\right]^2\right)+\right.\cr
&\ \ \ \ \ \ \ \ \ \ \ \ \ \ \ \ \ \ \ \ \ \ \left.+{1\over 9}
\left(1-3\left[{\vk_2.\vk_3\over k_2 k_3}\right]^2
+2 {\vk_1.\vk_2\ \vk_2.\vk_3\ \vk_3.\vk_1\over k_1^2 k_2^2 k_3^2}\right)
\right].\cr}\eqno(D.15)$$
If the initial density field has Gaussian fluctuations as described
in the part 2 the random variables $j_{\vk}\un$ are Gaussian and
$$\jk\un=-{D(t)\over D(t_i)}\dta_{\vk},$$
since $\dta(t,\vq)=[1-J(t,\vq)/J(t,\vq)]\approx -J\un(t,\vq)$ at the
first order.

The ensemble average of the product  $V\trois \dta_L$
exhibits the expectation value of the product of four Gaussian random
variables,
$$\eqalign{
\mg \dta_{\vk} j\un_{\vk_1} j\un_{\vk_2} j\un_{\vk_3} \md&=
-\left[{D(t)\over D(t_i)}\right]^3
\mg\dta_{\vk}\dta_{\vk_1}\dta_{\vk_2}\dta_{\vk_3}\md\cr
&=-\left[{D(t)\over D(t_i)}\right]^3
\left[\mg\dta_{\vk}\dta_{\vk_1}\md\ \mg\dta_{\vk_2}\dta_{\vk_3}\md+\sym.\right]
\cr
&=-\left[{D(t)\over D(t_i)}\right]^3
\left[P(k)\delta_3(\vk+\vk_1)\ P(k_2) \delta_3(\vk_2+\vk_3)+\sym.\right]\cr}
\eqno(D.16)$$
where $\sym.$ stands for three other terms obtained by a circular permutation
of the indices. The last expression holds because the random Gaussian
variables describe the density field {\sl
of a randomly homogeneous and isotropic
universe} with the power spectrum $P(k)$.

The use of the equations (D.14-16) gives
$$\eqalign{
\mg V\trois\ \dta_L\md={4V_0\over 21}\left[{D(t)\over D(t_i)}\right]^4
\int& {\d^3 \vk \over (2 \pi)^{3}} \WTH^2(k\ R_0) P(k)
{\d^3 \vk' \over (2 \pi)^{3}} P(k')\cr
&\times\left[1-{\left(\vk'.(\vk+\vk')\right)^2\over k'^2
\vert\vk+\vk'\vert^2}\right]
\left[1-\left({\vk.\vk'\over k k'}\right)^2\right].\cr}\eqno(D.17)$$
This integral can be simplified since the integration over
the angular part between $\vk$ and $\vk'$ can be made separately.
We can define the function $f(\tau)$ by
$$f(\tau)={1\over2}
\int_{-1}^1 \d u\ {(1-u^2)^2\over 1+\tau^2+2\tau u},\eqno(D.18)$$
then
$$\mg V\trois\ \dta_L\md={4V_0\over 21}\left[{D(t)\over D(t_i)}\right]^4
\int {\d^3 \vk \over (2 \pi)^{3}} \WTH^2(k\ R_0) P(k)
\int {\d^3 \vk' \over (2 \pi)^{3}} P(k') f(k'/k).\eqno(D.19)$$
The calculation of $f(\tau)$ can be done analytically and the result reads
$$\eqalign{
f(\tau)&={1\over 4\tau}\left[-2 A^3+{10\over3} A+(1-A^2)^2
\ln {A+1\over A-1}\right]\cr
A&={\tau\over2}+{1\over 2\tau}.\cr}
\eqno(D.20)$$
The shape of the function $f(\tau)$ is given in Fig. 3. Note that
$f(\tau)\sim 5/8$ for $\tau\to0$ and $f(\tau)\sim 5/8\tau^2$ for
$\tau\to\infty$, so that the integral (D.19) converges only if
$P(k)\ll k^{-1}$ when $k\to\infty$.

\vskip .5 cm
\leftline{\sl D.3 Calculation of $\mg [V\deux(t)]^2\md$}

 The expression of $V\deux(t)$ derives from the results given in (D.11),
$$V\deux(t)=
{2\over 7}\left[\DoDi\right]^2
\int{\d^3\vk_1\over (2\pi)^{3/2}}\ {\d^3\vk_2\over (2\pi)^{3/2}}\
j\un_{\vk_1}\WTH(k_1\,R_0)\ j\un_{\vk_2}\WTH(k_2\,R_0)\ (1-u^2)\eqno(D.21)$$
where $u=\vk_1.\vk_2/k_1k_2$. As a result we have
$$\eqalign{
\mg [V\deux(t)]^2\md=&
2\ \left({2\over 7}\right)^2\left[\DoDi\right]^4\ \int
{\d^3\vk_1\over (2\pi)^{3}}\ {\d^3\vk_2\over (2\pi)^{3}}\
\WTH^2\left(\vert\vk_1+\vk_2\vert\,R_0\right)\cr
&\ \times P(k_1)\ P(k_2)\ (1-u^2)^2.\cr}\eqno(D.22)$$
This expression can be simplified by the change of variable
$$\eqalign{
\vk_+=\vk_1+\vk_2&;\cr
\vk_-=\vk_1-\vk_2&;\cr}\eqno(D.23)$$
that leads to
$$\eqalign{
\mg [V\deux(t)]^2\md=&
\left({4\over 7}\right)^2\left[\DoDi\right]^4\ \int
{\d^3\vk_+\over (2\pi)^{3}}\ {\d^3\vk_-\over (2\pi)^{3}}\
\WTH^2(k_+\,R_0)\ P(k_1)\ P(k_2)\ \cr
&\ \times\left[{(1-v^2)\over \left(k_+/k_-+k_-/k_+\right)^2-4v^2}\right]^2
\cr}\eqno(D.24)$$
where $v=\vk_+.\vk_-/k_+k_-$. In case of a power--law spectrum
with an index $n$ (eq. [37]) the result reads,
$$\eqalign{
\mg [V\deux(t)]^2&\md=
\left({4\over 7}\right)^2\ 4^{-n}\left[\DoDi\right]^4\ \int
{\d^3\vk_+\over (2\pi)^{3}}\ k_+^{5+2n}\
\WTH^2(k_+\,R_0)\ \cr
&\times\int_{-1}^{1}\d v\int_{-\infty}^{+\infty}
{\d r\ r^{2+n}\over (2\pi)^{2}}\
(1-v^2)^2\ \left[ \left(k_+/k_-+k_-/k_+\right)^2-4v^2 \right]^{n/2-2},
\cr}\eqno(D.25)$$
which has to be integrated numerically.

 \vfill\eject

\centerline{\bf REFERENCES}
{\parindent=0pt
\apjref Balian, R. \& Schaeffer, R. 1989; A \& A; 220; 1;
\apjref Bardeen, J. M., Bond, J. R., Kaiser, N., \& Zsalay, A. S. 1986;
ApJ; 304; 15;
\apjref Bernardeau, F. 1992; ApJ; 192; 1;
\prepref Blanchard, A., Valls-Gabaud, D. \& Mamon, G.A., 1992; A \& A,
in press;
\apjref Bond, J.R., Cole, S., Efstathiou, G., \& Kaiser, N. 1991; ApJ; 379;
440;
\apjref Bouchet, F., Juszkiewicz, R., Colombi, S., \& Pellat, R. 1992; ApJ;
394; L5;
\apjref Bower, R.G. 1991; MNRAS; 248; 332;
\apjref Buchert, T. 1992; MNRAS; 254; 729;
\apjref Davis, M., Efstathiou, G., Frenk, C. \& White, S.D.M. 1985; ApJ;
292; 371;
\apjref Efstathiou, G. \& Rees, M.J. 1988; MNRAS; 230; 5p;
\apjref Evrard, A.E., \& Crone, M.M. 1992; ApJ; 394; L1;
\apjref Fisher, K.B., Davis, M., Strauss, M.A., Yahil, A. \& Huchra, J.P. 1993;
ApJ; 402; 42;
\apjref Fry, J. 1986; ApJ; 306; 358;
\apjref Henry, J.P. \& Arnaud, K.A. 1991; ApJ; 372; 410;
\prepref Juszkiewicz, R. \& Bouchet, F.; to appear in the proc.
of $2^{nd}$ DAEC workshop, Meudon observatory, March 1991, Ed. G. Mamon;
\apjref Lahav, O., Lilje, P.B., Primack, J.R. \& Rees, M.J. 1991; M.R.A.S.;
251; 128;
\apjref Kaiser, N. 1984; ApJ; 628; L9;
\prepref Katz, N., Quinn, T. \& Gelb, J.M. 1992; Fermilab preprint;
Kofman, L. 1991, in {\sl Primordial Nucleosynthesis \& Evolution of Early
Universe}, eds. Sato, K. \& Audouze, J., (Dordrecht: Kluwer)
\apjref Lilje, P.B. \& Lahav, O. 1991; ApJ; 347; 29;
\apjref Little, B., Weinberg, D.H., \& Park, C. 1991; MNRAS; 253; 307;
\apjref Moutarde, F, Alimi J.-M., Bouchet F.R., Pellat, R., Ramani,
A. 1991; ApJ; 382; 377;
\apjref Oukbir, J. \& Blanchard, A. 1992; ApJ; 262; L21;
\apjref Peacock, J.A. 1991; MNRAS; 253; 1p;
\bookref Peebles, P.J.E. 1980;  The Large Scale Structure of the Universe;
Princeton University Press, Princeton, N.J., USA;
\apjref Peebles, P.J.E. 1990; ApJ; 365; 27;
\apjref Peebles, P.J.E., Daly, R.A. \& Juszkiewicz, R. 1989; ApJ; 347; 563;
\apjref Press, W.H. \& Schechter, P. 1974; ApJ; 28; 19;
\apjref Thomas, P.A. \& Couchman, H.M.P. 1992; MNRAS; 257; 11;
\apjref Zel'dovich, Ya.B. 1970; A \& A; 5; 84;
}

 \vfill\eject

\centerline{\bf FIGURE CAPTIONS}
\noindent
\vskip .5cm
{\it Fig. 1 :}
Evolution of the density profile for an Einstein--de Sitter universe.
The initial profile is the
expected shape of the profile for various value of the power--law
index: $n=-2$ for (a), $n=-1$ for (b) and $n=0$ for (c).
The solid lines correspond to the initial profiles. The $n=0$
case corresponds to a Poisson distribution and the correlation length
is 0 which explains this top--hat shape of the initial profile.
The dashed lines
correspond to the exact nonlinear evolution of these profiles for various
time steps, $\dta_L=0.25, 0.5, 0.80, 1.10$.
\vskip .5 cm

{\it Fig. 2 :}
The nonlinear density as a function of $\dta_L$ (eq. [20]) for various
cosmological models. The solid line is the analytic expression
$\dta=(1-\dta_L/1.5)^{-1.5}-1$ corresponding to $\Omega\to 0$,
$\Lambda=0$; the dotted line is for $\Omega=1$, $\Lambda=0$; and the
dashed line is for $\Omega=1$, $\lambda=\Lambda/3H_0^2=1$.

\vskip .5 cm

{\it Fig. 3 :}
The function $f(\tau)$ (eq. [D.20]) as a function of $\tau$.
\vskip .5 cm

{\it Fig. 4 :}
The corrective coefficient, $\Cexp$ (eq. [34])
as a function of $n$ for a power--law spectrum (a) and
as a function of the cutoff, $k_c$, for a truncated power--law spectrum
and for various values of $n$ (b). $R$ is the size of the perturbation.
\vskip .5 cm

{\it Fig. 5 :}
Evolution of the density of a fluctuation as a function of the initial
overdensity and of the time. The upper solid line corresponds
to the spherical model. the second solid line corresponds
to the spherical model solution truncated at the third order. The long dashed
and small dashed lines correspond to the expected
evolution of a fluctuation up to the third order characterized respectively
by $\nu=4$ and $\nu=2$. The corrective coefficients are
calculated for $n=-2$.
\vskip .5 cm

{\it Fig. 6 :}
The corrective coefficient as a function of scale in case of a CDM
spectrum. For the mass scale of the clusters the correction is still small,
so that the spherical collapse model is expected to be a good description
for the collapse of a cluster.
\vskip .5 cm

{\it Fig. 7 :}
The coefficient $\Cscat$ (eq. [46]) as a function of $n$ (solid line) and
the coefficient $\Cscat'$ (eq. [48]) (dashed line).
\vskip .5 cm

{\it Fig. 8 :}
Schematic representation of the situation obtained in the $n\approx-2$ case.
The expectation behavior (dashed line) is close to the spherical collapse model
(solid line) with possible fluctuations represented by the shaded area, the
departure from the spherical collapse and the fluctuations being of the same
order of magnitude.

\end